\documentclass[11pt,pra,groupedaddress,floatfix]{article}
\usepackage{jheppub}
\usepackage{graphicx}
\usepackage{bm}
\usepackage{amsmath}
\usepackage{amsfonts}
\usepackage{amssymb}
\usepackage{mathtools}
\usepackage{latexsym}
\usepackage{slashed}
\usepackage{wrapfig}
\usepackage{nccmath}
\usepackage{nicefrac}
\usepackage{subcaption}
\usepackage{longtable}
\usepackage{tensor}
\usepackage[]{todonotes}
\usepackage{tikz}
\usepackage{tikz-feynman}

\renewcommand{\v}[1]{\boldsymbol{{#1}}}
\renewcommand{\[}{\begin{equation}\begin{aligned}}
\renewcommand{\]}{\end{aligned}\end{equation}}

\def\W{\mathbb{W}}
\def\P{\mathbb{P}}
\def\J{\mathbb{J}}
\def\s{s}

\def\dd{\hat d}
\def\del{\hat \delta}

\def\df{d \Phi}
\def\wn{\bar}

\def\sect#1{sect.~\ref{#1}}

\def\qb{\bar q}
\def\Lexp{\biggl\langle\!\!\!\biggl\langle}
\def\Rexp{\biggr\rangle\!\!\!\biggr\rangle}
\def\spinExp#1{\big\langle #1 \big\rangle}
\def\eqn{eq.}
\newcommand{\commentout}[1]{}

\abstract{
We develop a general formalism for computing classical observables for relativistic scattering of spinning particles, directly from on-shell amplitudes. We then apply this formalism to minimally coupled Einstein-gravity amplitudes for the scattering of massive spin 1/2 and spin 1 particles with a massive scalar, constructed using the double copy. In doing so we reproduce recent results at first post-Minkowskian order for the scattering of spinning black holes, through quadrupolar order in the spin-multipole expansion.
}

\begin{document}
	\title{Observables and amplitudes for spinning particles and black holes}
	\author[a]{Ben Maybee,}
	\author[a]{Donal O'Connell}
	\author[b]{and Justin Vines}
	\affiliation[a]{Higgs Centre for Theoretical Physics, School of Physics and Astronomy, The University of Edinburgh, Edinburgh, EH9 3JZ, Scotland, UK}
	\affiliation[b]{Max Planck Institute for Gravitational Physics (Albert Einstein Institute), Am M\"{u}hlenberg 1, Potsdam 14476, Germany}
	\emailAdd{b.maybee@ed.ac.uk}
	\emailAdd{donal@staffmail.ed.ac.uk}
	\emailAdd{justin.vines@aei.mpg.de}
	\emailAdd{}
	\maketitle
	\flushbottom

\section{Introduction}
\label{sec:intro}

Kerr black holes are very special spinning objects.  Any stationary axisymmetric extended body has an infinite tower of mass-multipole moments $\mathcal{I}_\ell$ and current-multipole moments $\mathcal{J}_\ell$, which generally depend intricately on its internal structure and composition.  For a Kerr black hole, every multipole is determined by only the mass $m$ and spin $s$, through the simple relation due to Hansen \cite{Hansen1974},
\begin{equation}
\mathcal I_\ell+i\mathcal J_\ell = m\left(\frac{is}{m}\right)^\ell \,.\label{eqn:multipoles}
\end{equation}
This distinctive behaviour is a reflection of the no-hair theorem, stating that black holes in general relativity (GR) are uniquely characterised by their mass and spin (and charge). 

Recent work has suggested that an on-shell expression of the no-hair theorem is that black holes correspond to minimal coupling in classical limits of quantum scattering amplitudes for massive spin $n$ particles and gravitons.  Amplitudes for long-range gravitational scattering of spin $\nicefrac{1}{2}$ and spin 1 particles were found in \cite{Ross:2007zza,Holstein:2008sx} to give the universal spin-orbit (pole-dipole level) couplings in the post-Newtonian corrections to the gravitational potential.  Further similar work in \cite{Vaidya:2014kza}, up to spin 2, suggested that the black-hole multipoles \eqref{eqn:multipoles} up to order $\ell=2n$ are faithfully reproduced from tree-level amplitudes for minimally coupled spin $n$ particles.

Such amplitudes for arbitrary spin $n$ were computed in \cite{Guevara:2017csg}, using the representation of minimal coupling for arbitrary spins presented in \cite{Arkani-Hamed:2017jhn} using the massive spinor-helicity formalism---see also~\cite{Conde:2016vxs,Conde:2016izb}.
Those amplitudes were shown in \cite{Guevara:2018wpp,Bautista:2019tdr} to lead in the limit $n\to\infty$ to the two-black-hole aligned-spin scattering angle found in \cite{Vines:2017hyw} at first post-Minkowskian (1PM) order and to all orders in the spin-multipole expansion, while in \cite{Chung:2018kqs} they were shown to yield the contributions to the interaction potential (for arbitrary spin orientations) at the leading post-Newtonian (PN) orders at each order in spin.  The importance of minimal coupling has been especially emphasised in \cite{Chung:2018kqs}, where, by matching at tree-level to the classical effective action of \cite{Levi:2015msa}, it was shown that the theory which reproduces the infinite-spin limit of minimally coupled graviton amplitudes is an effective field theory (EFT) of spinning black holes, with any deviation from minimal coupling adding further internal structure to the effective theory.

In this paper we remove the restriction to the aligned-spin configuration in the final results of \cite{Guevara:2018wpp,Bautista:2019tdr}, and the restriction to the nonrelativistic limit in the final results of \cite{Chung:2018kqs}.  We use on-shell amplitudes to directly compute relativistic classical observables for generic spinning-particle scattering, reproducing such results for black holes obtained by classical methods in \cite{Vines:2017hyw}, thereby providing more complete evidence for the correspondence between minimal coupling to gravity and classical black holes.



The dynamics of spinning black holes is of great interest for gravitational-wave astronomy \cite{Abbott:2016blz}, and spin leads to essential corrections which are required for precision analysis of signals from binary black hole mergers \cite{Buonanno:2014aza}.  Incorporating spin into a major theoretical platform for these experiments, the effective one-body formalism \cite{Buonanno:1998gg,Buonanno:2000ef}, is well established in the PN approximation \cite{Damour:2001tu,Damour:2008qf,Barausse:2009aa,Barausse:2009xi,Barausse:2011ys,Damour:2014sva,Bini:2017wfr}, and was recently extended to the PM approximation by means of a gauge-invariant spin holonomy \cite{Bini:2017xzy}. This has been used to compute the dipole (or spin-orbit) contribution to the conservative potential for two spinning bodies through 2PM order \cite{Bini:2018ywr}.
In addition to the all-multipole binary-black-hole results for generic spins at 1PM order found in \cite{Vines:2017hyw}, aligned-spin black-hole scattering has been considered also at 2PM order for low multipoles in \cite{Vines:2018gqi}. Meanwhile, calculating higher-order PN spin corrections has been a particular strength of the EFT treatment of PN dynamics \cite{Goldberger:2004jt,Porto:2005ac,Porto:2006bt,Porto:2008tb,Levi:2011eq,Levi:2015msa,Levi:2016ofk}; for reviews see \cite{Porto:2016pyg,Levi:2018nxp}.  All-multipole-order expressions are also known at the leading PN orders \cite{Vines:2016qwa,Siemonsen:2017yux}.

Alternatively, there exist many techniques for using scattering amplitudes to compute classical and quantum corrections to gravitational potentials \cite{Iwasaki:1971,Duff:1973zz,Donoghue:1993eb,Donoghue:1994dn,Bjerrum-Bohr:2002kt,Khriplovich:2004cx,Holstein:2004dn,Neill:2013wsa,Bjerrum-Bohr:2013bxa,Bjerrum-Bohr:2014lea,Bjerrum-Bohr:2014zsa,Bjerrum-Bohr:2016hpa,Bjerrum-Bohr:2017dxw,Cachazo:2017jef}, and there has been significant recent progress in obtaining such corrections in the PM approximation \cite{Damour:2016gwp,Damour:2017zjx,Bjerrum-Bohr:2018xdl,Cheung:2018wkq,Bern:2019nnu,Cristofoli:2019neg}.  Using amplitudes allows one to draw upon a powerful armoury of modern on-shell methods. Of particular applicability for gravity is the double copy \cite{Bern:2008qj,Bern:2010ue}, which asserts that replacing the colour factors in Yang-Mills amplitudes with kinematic factors satisfying the same Lie algebraic structure yields a gravity amplitude. While only proven at tree-level for pure gauge theory \cite{Bern:2010yg}, this conjecture can be applied to both massless and massive states \cite{Johansson:2014zca,Johansson:2015oia} and has a wealth of non-trivial supporting evidence \cite{Carrasco:2015iwa,Bern:2018jmv,Bern:2019isl}. It raises the provocative question of whether exact solutions in general relativity satisfy similar simple relationships to their classical Yang-Mills counterparts, coined the classical double copy \cite{Monteiro:2014cda,Luna:2015paa,Luna:2016due,Goldberger:2016iau,Luna:2016hge,Adamo:2017nia,Goldberger:2017frp,Bahjat-Abbas:2017htu,Carrillo-Gonzalez:2017iyj,Goldberger:2017vcg,Shen:2018ebu,Plefka:2018dpa,Berman:2018hwd,CarrilloGonzalez:2019gof,Luna:2018dpt}. Such a relationship has indeed been found in the gravitational radiation emitted by spinning sources at 1PM order \cite{Goldberger:2017ogt,Li:2018qap}. Irrespectively, amplitudes techniques have already been applied in general relativity to determine the sought-after 3PM correction to the conservative gravitational potential for the first time \cite{Bern:2019nnu} (see also \cite{Antonelli:2019ytb}), by matching to an effective theory of non-relativistic scalars \cite{Cheung:2018wkq}.

Calculating the gravitational potential is versatile but gauge-dependent. Amplitudes and observables, however, are on-shell and gauge-invariant, leading Kosower and two of the authors to introduce a direct mapping between the two \cite{Kosower:2018adc}. General formulae valid for massive scalar scattering in any quantum field theory, with interactions mediated by massless bosons, were written down for the impulse $\Delta p^\mu$ and total radiated momentum $R^\mu$, for any two-body scattering event. By analysing appropriate wavepackets and extracting powers of $\hbar$, results for these quantities in classical electrodynamics at tree and 1-loop levels were accurately reproduced. The relevance of the same classical limit for radiative scattering of massive scalars in Einstein gravity was also shown in \cite{Luna:2017dtq}.

In this paper we relax the restriction to scalars and consider conservative scattering of massive particles with spin. In addition to the (linear) impulse $\Delta p^\mu$, there is another relevant on-shell observable, the change $\Delta s^\mu$ in the spin (psuedo-)vector $s^\mu$, which we will call the angular impulse.  We introduce this quantity in \sect{sec:GRspin}, where we also review classical results from \cite{Vines:2017hyw} for binary black hole scattering at 1PM order.
In \sect{sec:QFTspin} we consider the quantum analogue of the spin vector, the Pauli-Lubanski operator; manipulations of this operator allow us to write expressions for the angular impulse akin to those for the linear impulse in \cite{Kosower:2018adc}. Obtaining the classical limit requires some care, which we discuss before constructing example gravity amplitudes in \sect{sec:amplitudes} from the double copy. In \sect{sec:KerrCalcs} we then show that substituting these examples into our general formalism exactly reproduces the leading terms of all-multipole order expressions for the impulse and angular impulse of spinning black holes \cite{Vines:2017hyw}. Finally, we discuss how our results further connect spinning black holes and scattering amplitudes in \sect{sec:discussion}.

\section{Spin and scattering observables in classical gravity}\label{sec:GRspin}

Before setting up our formalism for computing the angular impulse, let us briefly review aspects of this observable in relativistic classical physics. 

\subsection{Linear and angular momenta in asymptotic Minkowski space}

To describe the incoming and outgoing states for a weak scattering process in asymptotically flat spacetime, we can use special relativistic physics, working as in Minkowski spacetime. There, any isolated body has a constant linear momentum vector $p^\mu$ and an antisymmetric tensor field $J^{\mu\nu}(x)$ giving its total angular momentum about the point $x$, with the $x$-dependence determined by $J^{\mu\nu}(x')=J^{\mu\nu}(x)+2p^{[\mu}(x'-x)^{\nu]}$, or equivalently $\nabla_\lambda J^{\mu\nu}=2p^{[\mu}\delta^{\nu]}{}_\lambda$.

Relativistically, center-of-mass (cm) position and intrinsic and orbital angular momenta are frame-dependent concepts, but a natural inertial frame is provided by the direction of the momentum $p^\mu$, giving the proper rest frame.  We define the body's proper cm worldline to be the set of points $z$ such that $J^{\mu\nu}(z)p_\nu=0$, i.e.\ the proper-rest-frame mass-dipole vector about $z$ vanishes, and we can then write
\begin{equation}\label{eqn:Jmunu}
J^{\mu\nu}(x)=2p^{[\mu}(x-z)^{\nu]}+S^{\mu\nu},
\end{equation}
where $z$ can be any point on the proper cm worldline, and where $S^{\mu\nu}=J^{\mu\nu}(z)$ is the intrinsic spin tensor, satisfying
\begin{equation}
S^{\mu\nu} p_\nu=0.\label{eqn:SSC}
\end{equation}
Equation \eqref{eqn:SSC} is often called the ``covariant'' or Tulczyjew-Dixon spin supplementary condition (SSC) \cite{Fokker:1929,Tulczyjew:1959} in its (direct) generalization to curved spacetime in the context of the Mathisson-Papapetrou-Dixon equations \cite{Mathisson:1937zz,Mathisson:2010,Papapetrou:1951pa,Dixon1979,Dixon:2015vxa} for the motion of spinning extended test bodies.
Given the condition \eqref{eqn:SSC}, the complete information of the spin tensor $S^{\mu\nu}$ is encoded in the momentum $p^\mu$ and the spin pseudo-vector \cite{Weinberg:1972kfs},
\begin{equation}
s_\mu = \frac{1}{2m}\epsilon_{\mu\nu\rho\sigma} p^\nu S^{\rho\sigma} = \frac{1}{2m}\epsilon_{\mu\nu\rho\sigma} p^\nu J^{\rho\sigma}(x),\label{eqn:GRspinVec}
\end{equation}
where $\epsilon_{0123} = +1$ and the metric signature is mostly minus, with $p^2=m^2$.  Note that $s\cdot p=0$; $s^\mu$ is a spatial vector in the proper rest frame.
Given \eqref{eqn:SSC}, the inversion of the first equality of \eqref{eqn:GRspinVec} is
\begin{equation}
S_{\mu\nu} =\frac{1}{m} \epsilon_{\mu\nu\lambda\tau} p^\lambda s^\tau.\label{eqn:SSCintrinsicSpin}
\end{equation}
The total angular momentum tensor $J^{\mu\nu}(x)$ can be reconstructed from $p^\mu$, $s^\mu$, and a point $z$ on the proper cm worldline, via \eqref{eqn:SSCintrinsicSpin} and \eqref{eqn:Jmunu}.

\subsection{Scattering of spinning black holes in linearized gravity}

Following the no-hair property emphasised in \sect{sec:intro}, the full tower of gravitational multipole moments of a spinning black hole, and thus also its (linearized) gravitational field, are uniquely determined by its monopole $p^\mu$ and dipole $J^{\mu\nu}$.  This is reflected in the scattering of two spinning black holes, in that the net changes in the holes' linear and angular momenta depend only on their incoming linear and angular momenta.  It has been argued in \cite{Vines:2017hyw} that the following results concerning two-spinning-black-hole scattering, in the  1PM approximation to GR, follow from the linearized Einstein equation and a minimal effective action description of spinning black hole motion, the form of which is uniquely fixed at 1PM order by general covariance and appropriate matching to the Kerr solution.



Consider two black holes with incoming momenta $p_1^\mu=m_1 u_1^\mu$ and $p_2^\mu=m_2 u_2^\mu$, defining the 4-velocities $u^\mu=p^\mu/m$ with $u^2=1$, and incoming spin vectors $s_1^\mu=m_1 a_1^\mu$ and $s_2^\mu=m_2 a_2^\mu$, defining the rescaled spins $a^\mu=s^\mu/m$ (with units of length, whose magnitudes measure the radii of the ring singularities).  Say the holes' zeroth-order incoming proper cm worldlines are orthogonally separated at closest approach by a vectorial impact parameter $b^\mu$, pointing from 2 to 1, with $b\cdot u_1 =b\cdot u_2=0$.  Then, according to the analysis of \cite{Vines:2017hyw}, the net changes in the momentum and spin vectors of black hole 1 are given by
\begin{alignat}{3}
\begin{aligned}
\Delta p_1^{\mu} &= \textrm{Re}\{\mathcal Z^\mu\}+O(G^2),
\\
\Delta s_1^{\mu} &= - u_1^\mu a_1^\nu\, \textrm{Re}\{\mathcal Z_\nu\} - \epsilon^{\mu\nu\alpha\beta} u_{1\alpha} a_{1\beta}\, \textrm{Im}\{\mathcal Z_\nu\}+O(G^2),
\end{aligned}\label{eqn:KerrDeflections}
\end{alignat}
where
\begin{equation}
\mathcal Z_\mu = \frac{2G m_1 m_2}{\sqrt{\gamma^2 - 1}}\Big[(2\gamma^2 - 1)\eta_{\mu\nu} - 2i\gamma \epsilon_{\mu\nu\alpha\beta} u_1^\alpha u_2^\beta\Big]\frac{ b^\nu + i\Pi^\nu{ }_\rho (a_1+a_2)^\rho}{[b + i\Pi(a_1+a_2)]^2}\,,
\end{equation}
with $\gamma = u_1\cdot u_2$ being the relative Lorentz factor, and with
\begin{equation}
\begin{aligned}
\Pi^\mu{ }_\nu &= \epsilon^{\mu\rho\alpha\beta} \epsilon_{\nu\rho\gamma\delta} \frac{{u_1}_\alpha {u_2}_\beta u_1^\gamma u_2^\delta}{\gamma^2 - 1}\\ &= \delta^\mu{ }_\nu +\frac1{\gamma^2 - 1}\bigg(u_1^\mu({u_1}_\nu - \gamma {u_2}_\nu) + u_2^\mu({u_2}_\nu - \gamma {u_1}_\nu)\bigg) \label{eqn:projector}
\end{aligned}
\end{equation}
being the projector into the plane orthogonal to both incoming velocities.
The analogous results for black hole 2 are given by interchanging the identities $1\leftrightarrow 2$.

If we take black hole 2 to have zero spin, $a_2^\mu\to0$, and if we expand to quadratic order in the spin of black hole 1, corresponding to the quadrupole level in 1's multipole expansion, then we obtain the results shown in \eqref{eqn:JustinImpResult} and \eqref{eqn:JustinSpinResult} below.  In the remainder of this paper, developing necessary tools along the way, we show how those results can be obtained from classical limits of scattering amplitudes for one-graviton exchange between a massive scalar particle and a massive spin-$n$ particle, with minimal coupling to gravity, with $n=1/2$ to yield the dipole level, and with $n=1$ to yield the quadrupole level.

\section{Spin and scattering observables in quantum field theory}
\label{sec:QFTspin}

The linear and angular impulses, $\Delta p^\mu$ and $\Delta s^\mu$, are observable, on-shell quantities. In \cite{Kosower:2018adc} a general formalism for calculating the classical impulse $\Delta p^\mu$ in quantum field theory was introduced; as the angular impulse is also on-shell similar methods should be applicable. A first task is to understand what quantum mechanical quantity corresponds to the classical spin pseudovector of equation~\eqref{eqn:GRspinVec}. This spin vector is a quantity associated with a single classical body, and we therefore begin by discussing single particle states (to set up our notation) before discussing the spin vector of a quantum state. We then move on to the change in spin during a scattering event, and finally we will explain the correspondence region in which a quantum calculation must agree with a classical one. As our aim is to address black hole scattering processes, we restrict throughout to the case in which our incoming and outgoing particles are massive.

\subsection{Single particle states}

We will be interested in both bosonic and fermionic particles, normalising creation and annihilation operators so that
\begin{equation}
[a_i(p), a_j^\dagger(q)]_\pm = \del_\Phi(p - q) \delta_{ij}\,,
\end{equation}
where $\del_\Phi(p)$ is the appropriate delta function for the on-shell phase-space measure:
\[
\del_\Phi(p) \equiv 2E_p (2\pi)^3 \delta^{(3)}(\v{ p}).
\]
Single particle states of a given momentum and spin are defined, as usual, by $|p, i\rangle = a^\dagger_i(p) |0\rangle$. Notice that the index $i$ transforms under the little group, which for a massive particle in four dimensions is SU(2).

Our interest will primarily be in spatially localised particles, which are associated with a wavefunction $\phi(p)$ in momentum space. In general there is also a little group index on the wavefunction; for our purposes it is sufficient to consider wavefunctions of the form $\phi(p) \xi_i$. Thus, we will concern ourselves with states of the form
\begin{equation}
|\psi\rangle = \sum_{i}\int \df(p)\, \phi(p) \xi_i\! \left|p, i\right\rangle\,,
\label{eq:singleParticleState}
\end{equation}
where the invariant phase space measure is
\[
\df(k) = \dd^4k\, \del^{(+)}(k^2 - m^2) \equiv \frac{d^4k}{(2\pi)^4}\, 2\pi \Theta(k^0) \delta(k^2 - m^2)\
\]
We normalise the wavefunction by choosing $\int \df(p) |\phi(p)|^2 = \sum_i|\xi_i|^2 = 1$. 

\subsection{The Pauli-Lubanski spin pseudovector}

Now we turn to the question of what operator in quantum field theory is related to the classical spin pseudovector of equation~\eqref{eqn:GRspinVec}. We propose that the correct quantum-mechanical interpretation is that the spin is nothing but the expectation value of the Pauli-Lubanski operator
\begin{equation}
\W_\mu = \frac{1}{2}\epsilon_{\mu\nu\rho\sigma} \P^\nu \J^{\rho\sigma}\,,\label{eqn:PLvec}
\end{equation}
where $\P^\mu$ and $\J^{\rho\sigma}$ are the translation and Lorentz generators, respectively. In particular, our claim is that the expectation value
\begin{equation}
\langle s^\mu \rangle \equiv \frac1{m}\langle \mathbb{W}^\mu \rangle = \frac{1}{2m} \epsilon_{\mu\nu\rho\sigma}\langle\P^\nu \J^{\rho\sigma}\rangle  \label{eqn:defOfQFTspinVec}
\end{equation}
of the Pauli-Lubanski operator on a single particle state~\eqref{eq:singleParticleState} is the quantum-mechanical generalisation of the classical spin pseudo-vector. Indeed a simple comparison of equations~\eqref{eqn:GRspinVec} and~\eqref{eqn:PLvec} indicates a connection between the two quantities. We will provide abundant evidence for this link in the remainder of this article. Matrix elements of the Pauli-Lubanski vector are also relevant in the context of hadronic physics~\cite{Cotogno:2019xcl,Lorce:2019sbq}.

The Pauli-Lubanski operator is a basic quantity in the classification of free particle states, although it receives less attention in introductory accounts of quantum field theory than it should. With the help of the Lorentz algebra
\[
[\J^{\mu\nu}, \P^\rho] &= i \hbar (\eta^{\mu\rho} \P^\nu - \eta^{\nu\rho} \P^\mu) \,, \\
[\J^{\mu\nu}, \J^{\rho\sigma}] &= i \hbar (\eta^{\nu\rho} \J^{\mu\sigma} - \eta^{\mu\rho} \J^{\nu\sigma} - \eta^{\nu\sigma} \J^{\mu\rho} + \eta^{\mu\rho} \J^{\mu\sigma}) \, ,
\]
it is easy to establish the important fact that the Pauli-Lubanski operator commutes with the momentum:
\[
[\P^\mu, \W^\nu] = 0.
\]
Furthermore, as $\W^\mu$ is a vector operator, it satisfies
\[
[\J^{\mu\nu}, \W^\rho] = i\hbar (\eta^{\mu\rho} \W^\nu - \eta^{\nu\rho} \W^\mu) \,.
\]
It then follows that the commutation relations of $\W$ with itself are
\[
[\W^\mu, \W^\nu] = -i\hbar \epsilon^{\mu\nu\rho\sigma} \W_\rho \P_\sigma.
\]
On single particle states this last commutation relation takes a particularly instructive form. Working in the rest frame of our massive particle state, evidently $W^0 = 0$. The remaining generators satisfy\footnote{We normalise $\varepsilon^{123} = +1$, as usual.}
\[
[ \W^i, \W^j] = i \hbar \varepsilon^{ijk} \W^k \,,
\]
so that the Pauli-Lubanski operators are nothing but the generators of the little group. Not only is this the basis for their importance, but also we will find that these commutation relations are directly useful in our computation of the change in a particle's spin during scattering.

Because $\W^\mu$ commutes with the momentum, we have
\[
\langle p', j| \W^\mu |p, i \rangle \propto \del_\Phi(p-p').\label{eqn:PLdeltaFunc}
\]
We define the matrix elements of $\W$ on the states of a given momentum to be
\[
\langle p', j| \W^\mu |p, i \rangle \equiv m \s^\mu_{ij}(p)\, \del_\Phi(p-p') \,, \label{eqn:PLinnerProd}
\]
so that the expectation value of the spin vector is
\[
\langle \s^\mu \rangle = \int d\Phi(p) \, | \phi(p) |^2 \, \xi^*_i \s^\mu_{ij} \xi_j.
\]
The matrix $\s^\mu_{ij}(p)$, sometimes called the spin polarisation vector, will be important below. These matrices inherit the commutation relations of the Pauli-Lubanski vector, so that in particular
\[
[\s^\mu(p), \s^\nu(p) ] = - i \frac{\hbar}{m} \, \epsilon^{\mu\nu\rho\sigma} \s_\rho(p) p_\sigma \,.
\]

Specialising now to a particle in a given representation, we may derive well-known \cite{Ross:2007zza,Holstein:2008sx,Bjerrum-Bohr:2013bxa,Bjerrum-Bohr:2014lea,Bjerrum-Bohr:2017dxw,Guevara:2017csg} explicit expressions for the spin polarisation $\s^\mu_{ij}(p)$ starting from the Noether current associated with angular momentum. We provide details in Appendix~\ref{sec:spinVecDetails} for the simple spin 1/2 and 1 cases. For a Dirac spin $1/2$ particle, the spin polarisation is
\[
s^\mu_{ab}(p) = \frac{\hbar}{4m} \bar{u}_a(p) \gamma^\mu \gamma^5 u_b(p)\,.\label{eqn:spinorSpinVec}
\]
Meanwhile, for massive vector bosons we have
\begin{equation}
s_{ij}^\mu(p) = \frac{i\hbar}{m} \epsilon^{\mu\nu\rho\sigma} p_\nu \varepsilon{^*_i}_\rho(p) {\varepsilon_j}_\sigma(p)\,. \label{eqn:vectorSpinVec}
\end{equation}
We have normalised these quantities consistent with the algebraic properties of the Pauli-Lubanski operator.

\subsection{The change in spin during scattering}

Now that we have a quantum-mechanical understanding of the spin vector, we move on to discuss the dynamics of the spin vector in a scattering process. Following the set-up in~\cite{Kosower:2018adc} we consider the scattering of two stable, massive particles which are quanta of different fields, and are separated by an impact parameter $b^\mu$. We will explicitly consider scattering processes mediated by vector bosons and gravitons.
The relevant incoming two-particle state is
\begin{equation}
|\Psi\rangle = \sum_{a_1, a_2}\int \df(p_1) \df(p_2)\,\phi_1(p_1) \phi_2(p_2) \xi_{a_1} \xi_{a_2} e^{ib\cdot p_1/\hbar}\left|p_1\, p_2; a_1\, a_2\right\rangle\,,\label{eqn:inState}
\end{equation}
where the displacement operator insertion accounts for the particles' spatial separation. 

The initial spin vector of particle 1 is
\[
\langle s_1^\mu \rangle = \frac1{m_1} \langle \Psi |\W^\mu_1 |\psi \rangle\,,
\]
where $\W^\mu_1$ is the Pauli-Lubanski operator of the field corresponding to particle 1. Since the $S$ matrix is the time evolution operator from the far past to the far future, the final spin vector of particle 1 is
\[
\langle s_1'^\mu \rangle = \frac1{m_1} \langle \Psi | S^\dagger \W^\mu_1 S| \psi \rangle\,.
\]
We define the angular impulse on particle 1 as the difference between these quantities:
\begin{equation}
\langle \Delta s_1^\mu \rangle = \frac1{m_1}\langle\Psi|S^\dagger \mathbb{W}_1^\mu S|\Psi\rangle - \frac1{m_1}\langle\Psi|\mathbb{W}_1^\mu|\Psi\rangle\,.\label{eqn:defOfAngImp}
\end{equation}
Writing $S = 1 + iT$ and making use of the optical theorem yields
\begin{equation}
\langle \Delta s_1^\mu\rangle = \frac{i}{m_1}\langle\Psi|[\mathbb{W}_1^\mu,{T}]|\Psi\rangle + \frac{1}{m_1}\langle\Psi|{T}^\dagger[\mathbb{W}_1^\mu,{T}]|\Psi\rangle\,.\label{eqn:spinShift}
\end{equation}
It is clear that the second of these terms will lead to twice as many powers of the coupling constant for a given interaction. Therefore only the first term is able to contribute at leading order. In this paper we exclusively consider tree level scattering $\mathcal{A}^{(0)}$, so the first term is the sole focus of our attention.

Our goal now is to express the leading-order angular impulse in terms of amplitudes. To that end we substitute the incoming state in equation~\eqref{eqn:inState} into the first term of \eqn~\eqref{eqn:spinShift}, and the leading-order angular impulse is given by
\begin{multline}
\langle\Delta s^{\mu,(0)}_1\rangle = \frac{i}{m_1}\sum_{a'_1, a_1} \sum_{a'_2, a_2} \int \df(p_1') \df(p_2') \df(p_1) \df(p_2)\,\phi_1^*(p_1') \phi_2^*(p_2') \phi_1(p_1) \phi_2(p_2) \\ \times {\xi_1}^*_{a'_1} {\xi_2}^*_{a'_2} {\xi_1}_{a_1} {\xi_2}_{a_2} e^{ib\cdot(p_1-p'_1)/\hbar} \left\langle p_1'\,p_2';a'_1\, a'_2\left|\W^\mu\, {T} -{T}\, \W^\mu \right|p_1\,p_2; a_1\, a_2 \right\rangle.
\end{multline}
Scattering amplitudes can now be explicitly introduced by inserting a complete set of states
\begin{equation}
\mathbb{I} = \sum_{b_1,b_2} \int\! \df(r_1) \df(r_2)\, |r_1\, r_2; b_1\, b_2\rangle \langle r_1\, r_2; b_1\, b_2|
\end{equation}
between the spin and interaction operators. In their first appearance this yields
\begin{multline}
\sum_{b_1,b_2} \int\! \df(r_1) \df(r_2) \langle p'_1\, p'_2; a'_1\, a'_2|\W^\mu|r_1\, r_2;b_1\, b_2\rangle \langle r_1\, r_2;b_1\, b_2|T|p_1\,p_2; a_1\, a_2\rangle = \\
m_1 \sum_{b_1} \int\! \df(r_1) {\s}^\mu_{1\, a'_1b_1}(p'_1)\, \del_\Phi(p'_1 - r_1)\,
 \mathcal{A}_{b_1 a'_2 a_1 a_2}(p_1, p_2 \rightarrow r_1, p'_2) \del^{(4)}(r_1 + p'_2 - p_1 - p_2)\,,
\end{multline}
where, along with the definition of the scattering amplitude, we have used the definition of the spin polarisation vector~\eqref{eqn:spinorSpinVec}. The result for the other ordering of $T$ and $\W^\mu$ is very similar. We will suppress the summation over repeated spin indices from now on.

Substituting into the full expression for $\langle\Delta s_1^{\mu,(0)}\rangle$ and integrating over the delta functions, we find that the angular impulse is
\begin{equation}
\begin{aligned}
\langle\Delta s^{\mu,(0)}_1\rangle =& i \int \df(p_1') \df(p_2') \df(p_1) \df(p_2)\,\phi_1^*(p_1') \phi_2^*(p_2') \\ & \quad\times \phi_1(p_1) \phi_2(p_2) {\xi_1}^*_{a'_1} {\xi_2}^*_{a'_2} {\xi_1}_{a_1} {\xi_2}_{a_2} e^{ib\cdot(p_1-p'_1)/\hbar} \del^{(4)}(p_1' + p_2' - p_1 - p_2) \\ & \quad\times\bigg(\s_{1\, a'_1 b_1}^\mu(p_1') \mathcal{A}_{b_1 a'_2 a_1 a_2}(p_1, p_2 \rightarrow p'_1, p'_2) \\ &\qquad\qquad\qquad\qquad\qquad\quad - \mathcal{A}_{a'_1 a'_2 b_1 a_2}(p_1, p_2 \rightarrow p'_1, p'_2) \s_{1 \, b_1 a_1}^{\mu}(p_1)\bigg).
\end{aligned}
\end{equation}
We now eliminate the delta function by introducing the momentum mismatch $q_i = p'_i - p_i$ and performing an integral. The leading-order angular impulse becomes
\begin{equation}
\begin{aligned}
\langle\Delta s^{\mu,(0)}_1\rangle = i & \int \df(p_1) \df(p_2)\, \dd^4q\,\del(2p_1\cdot q + q^2) \del(2p_2\cdot q - q^2) \\ \times& \phi_1^*(p_1 + q) \phi_2^*(p_2 - q) \phi_1(p_1) \phi_2(p_2) {\xi_1}^*_{a'_1} {\xi_2}^*_{a'_2} {\xi_1}_{a_1} {\xi_2}_{a_2} e^{-ib\cdot q/\hbar} \\\times& \bigg(\s_{1\, a'_1 b_1}^{\mu}(p_1 + q) \mathcal{A}_{b_1 a'_2 a_1 a_2}(p_1, p_2 \rightarrow p_1 + q, p_2 - q) \\ &\qquad\qquad - \mathcal{A}_{a'_1 a'_2 b_1 a_2}(p_1, p_2 \rightarrow p_1 + q, p_2 - q) \s_{1\,b_1 a_1}^{\mu}(p_1)\bigg).\label{eqn:exactAngImp}
\end{aligned}
\end{equation}

\subsection{Passing to the classical limit}
\label{sec:classicalLim}

The previous expression is an exact, quantum formula for the change in the spin vector during conservative two-body scattering. As a well-defined observable, we can extract the classical limit of the angular impulse by following the formalism introduced in~\cite{Kosower:2018adc}, which contains a careful and covariant discussion of the correspondence regime in which the classical and quantum theories must agree. 

We limit ourselves to a simplified, intuitive version of this classical limit. The basic idea is simple: the wavefunctions must localise the particles, without leading to a large uncertainty in the momenta of the particles. They therefore have a finite but small width $\Delta x = \ell_w$ in position space, and $\Delta p = \hbar/\ell_w$ in momentum space. This narrow width restricts the range of the integral over $q$ in equation~\eqref{eqn:exactAngImp} so that $q \lesssim \hbar /\ell_w$. We therefore introduce the wavenumber $\qb = q/\hbar$.
We further assume that the wavefunctions are very sharply peaked in momentum space around the value $\langle p_i^\mu \rangle = m_i u_i^\mu$, where $u_i^\mu$ is a classical proper velocity. We neglect the small shift $q = \hbar \qb$ in the wavefunctions present in equation~\eqref{eqn:exactAngImp}, and also the term $q^2$ compared to the dominant $2 p \cdot q$ in the delta functions, arriving at
\[
\langle\Delta s^{\mu,(0)}_1\rangle = i & \int \df(p_1) \df(p_2)\, \dd^4q\,\del(2p_1\cdot q ) \del(2p_2\cdot q) |\phi_1(p_1)|^2 |\phi_2(p_2)|^2 e^{-ib\cdot q/\hbar} \\\times& {\xi_1}^*_{a'_1} {\xi_2}^*_{a'_2}\bigg(\s_{1\, a'_1 b_1}^{\mu}(p_1 + q) \mathcal{A}_{b_1 a'_2 a_1 a_2}(p_1, p_2 \rightarrow p_1 + q, p_2 - q) \\ &\qquad\qquad - \mathcal{A}_{a'_1 a'_2 b_1 a_2}(p_1, p_2 \rightarrow p_1 + q, p_2 - q) \s_{1\,b_1 a_1}^{\mu}(p_1)\bigg)  {\xi_1}_{a_1} {\xi_2}_{a_2} \,.
\]
It is convenient to introduce a notation for the expectation values over the wavefunctions
\begin{multline}
\Lexp f(p_1,p_2,\ldots) \Rexp \equiv \sum_{a'_1, a_1} \sum_{a'_2, a_2} \int \df(p_1)\df(p_2)\;|\phi_1(p_1)|^2\,|\phi_2(p_2)|^2\\ \times{\xi^*}_{1\, a'_1} {\xi^*}_{1\, a'_2} 
f^{a'_1 a'_2 a_1 a_2}(p_1,p_2,\ldots) \xi_{1\, a_1} \xi_{2\, a_2}\,, \label{eqn:doubleAngleBrackets}
\end{multline}
so that our angular impulse takes the form
\[
\langle\Delta s^{\mu,(0)}_1\rangle = i & \Lexp \int \dd^4 q \del (2p_1 \cdot q) \del (2p_2\cdot q) e^{-ib\cdot q/\hbar}  \bigg(\s^{\mu}(p_1 + \qb \hbar) \mathcal{A}(p_1, p_2 \rightarrow p_1 + q, p_2 - q) \\ &\qquad\qquad - \mathcal{A}(p_1, p_2 \rightarrow p_1 + q, p_2 - q) \s_{1}^{\mu}(p_1)\bigg)\Rexp.
\label{eq:intermediate}
\]
Notice that in equation~\eqref{eq:intermediate}, both the spin vector and the amplitude are matrices with spinor indices, some of which are contracted together.
Reference~\cite{Kosower:2018adc} presents a more careful and covariant treatment of this process.

An important $\hbar$ shift remaining is that of the spin polarisation vector $\s_{1\, a'_1b_1}^{\mu}(p_1 + \hbar\wn q)$. This object is a Lorentz boost of $\s_{1\, a'_1b_1}^{\mu}(p_1)$. In the classical limit $q$ is small, so the Lorentz boost $\Lambda^{\mu}{ }_\nu p_1^\nu = p_1^\mu + \hbar\wn q^\mu$ is infinitesimal.
In the vector representation an infinitesimal Lorentz transformation is $\Lambda^\mu\,_\nu=\delta^\mu_\nu + \omega^\mu{ }_\nu$, so for our boosted momenta $\omega^\mu{ }_\nu p_1^\nu = \hbar\wn q^\mu$. The appropriate generator is
\begin{equation}
\omega^{\mu\nu} = -\frac{\hbar}{m_1^2}\left(p_1^\mu \wn q^\nu - \wn q^\mu p_1^\nu\right)\,.\label{eqn:LorentzParameters}
\end{equation} 
This result is valid for particles of any spin as it is purely kinematic, and therefore can be universally applied in our general formula for the angular impulse. In particular, since $\omega_{\mu\nu}$ is explicitly $\mathcal{O}(\hbar)$ the spin polarisation vector transforms as
\begin{equation}
\s_{1\,ab}^{\mu}(p_1 + \hbar\wn q) = \s_{1\,ab}^{\mu}(p_1) - \frac\hbar{m^2} p^\mu \qb \cdot s_{ab}(p_1).
\label{eqn:infinitesimalBoost}
\end{equation}
The angular impulse becomes
\begin{multline}
\langle \Delta s_1^{\mu,(0)} \rangle \rightarrow \Delta s_1^{\mu,(0)}
\\
 = \Lexp i \int\!\dd^4\wn q\,\del(2p_1\cdot\wn q) \del(2p_2\cdot\wn q)e^{-ib\cdot\wn q} \bigg(-\hbar^3 \frac{p_1^\mu}{m_1^2} \qb \cdot \s_{1} (p_1) \mathcal{A}(q) + \hbar^2 \big[\s^\mu_1(p_1), \mathcal{A}(q)\big] \bigg)\Rexp.\label{eqn:limAngImp}
\end{multline}
The appearance of a commutator is a manifestation of the spin indices in \eqn~\eqref{eqn:exactAngImp}, which are left implicit under the double angle brackets. The formula appears to be of a non-uniform order in $\hbar$, but fortunately this is not really the case: any terms in the amplitude with diagonal indices will trivially vanish under the commutator; alternatively, any term with a commutator will introduce a factor of $\hbar$ through the algebra of the Pauli-Lubanski vectors.
Therefore all terms have the same weight, $\hbar^3$, independently of factors appearing in the amplitude. 
An analogous formula\footnote{Note we have modified the definition of the double angle brackets from \cite{Kosower:2018adc} by including spins. For scalar amplitudes these terms drop out as the amplitude spin structure must be diagonal.} for the leading order, classical, linear impulse is~\cite{Kosower:2018adc}
\begin{equation}
\Delta p_1^{\mu,(0)} = \Lexp i\hbar^3 \int\! \dd^4 \wn{q}\, \del(2p_1\cdot\wn q) \del(2p_2\cdot\wn q) e^{-ib\cdot\wn q}\, \wn q^\mu \mathcal{A}^{(0)}(p_1, p_2 \rightarrow p_1 + \hbar\wn q, p_2 - \hbar\wn q) \Rexp \, . \label{eqn:limImpulse}
\end{equation}
We will make use of both the linear and angular impulse formulae below.

There is a caveat regarding the uncertainty principle in the context of our spinning particles. In this article we restrict to low spins: spin 1/2 and spin 1. Consequently the expectation of the spin vector $\langle s^\mu \rangle$ is of order $\hbar$; indeed $\langle s^2 \rangle = n(n+1) \hbar^2$. This requires us to face the quantum-mechanical distinction between $\langle s^\mu s^\nu \rangle$ and $\langle s^\mu \rangle \langle s^\nu \rangle$. Because of the uncertainty principle, the uncertainty $\sigma_1^2$ associated with the operator $s^1$, for example, is of order $\hbar$, and therefore the difference between $\langle s_1^2 \rangle$ and $\langle s_1 \rangle^2$ is of order $\hbar^2$. Thus the difference $\langle s^\mu s^\nu \rangle - \langle s^\mu \rangle \langle s^\nu \rangle$ is of order $\langle s^\mu s^\nu \rangle$. We are therefore not entitled to replace $\langle s^\mu s^\nu \rangle$ by $\langle s^\mu \rangle \langle s^\nu \rangle$, and will make the distinction between these quantities below. One can overcome this limitation by studying very large spin representations, in which case a scaling limit is available to suppress $\langle s^\mu s^\nu \rangle - \langle s^\mu \rangle \langle s^\nu \rangle$.

The procedure for passing from amplitudes to a concrete expectation value is as follows. Once one has computed the amplitude, and evaluated any commutators, explicit powers of $\hbar$ must cancel. We then evaluate the integrals over the on-shell phase space of the incoming particles simply by evaluating the momenta $p_i$ as $p_i = m_i u_i$. An expectation value over the spin wave functions $\xi$ remains; these are always of the form $\langle s^{\mu_1} \cdots s^{\mu_n}\rangle$ for various values of $n$. 

\section{Classical limits of amplitudes with spin}
\label{sec:amplitudes}

We have constructed a general formula for calculating the leading classical contribution to the angular impulse from scattering amplitudes. In the limit these amplitudes are Laurent expanded in $\hbar$, with only one term in the expansion providing a non-zero contribution. How this expansion works in the case for scalar amplitudes was established in \cite{Kosower:2018adc,Luna:2017dtq}, but now we need to consider examples of amplitudes for particles with spin. The identification of the spin polarisation vector defined in \eqn~\eqref{eqn:PLinnerProd} will be crucial to this limit.

We will again look at the two lowest spin cases, considering tree level scattering of a spin $\nicefrac{1}{2}$ or spin $1$ particle off a scalar in Yang-Mills theory and gravity. Yang-Mills amplitudes will be denoted by $\mathcal{A}_{n_1-0}$, and those for Einstein gravity as $\mathcal{M}_{n_1-0}$.

\subsection{Gauge theory amplitudes}

Our gauge theory consists of Yang-Mills theory minimally coupled to matter in the fundamental representation of the gauge group. The common Lagrangian will be
\begin{equation}
\mathcal{L}_0 = -\frac{1}{2} {\rm tr}F_{\mu\nu}^a F^{\mu\nu}_a + \sum_i\left[(D_\mu\Phi_i)^\dagger (D^\mu\Phi_i) - m_i^2|\Phi_i|^2 \right],\label{eqn:scalarQCDLagrangian}
\end{equation}
with coupling constant $\tilde{g} = g/\sqrt{\hbar}$ and $\Phi_i$ two massive scalars \cite{Luna:2017dtq}. For the amplitudes relevant for our on-shell observables only the $t$ channel contributes, so with colour factors $\tilde{T}^a$ = $\sqrt{2} T^a$\footnote{We choose this normalisation as it simplifies the colour replacements in the double copy.} the full tree-level amplitude is
\begin{equation}
i\mathcal{A}_{0-0} = \frac{i\tilde{g}^2}{2q^2}(2p_1+q)\cdot(2p_2-q)\, \tilde{T}_1\cdot \tilde{T}_2\,.
\end{equation}
The only classically significant contribution from this amplitude comes from the leading order term in the $\hbar$ Laurent expansion. Factors of $\hbar$ enter the amplitude in the coupling constant and from the restriction $q = \hbar\wn q$, yielding
\begin{equation}
\mathcal{A}_{0-0} = \frac{g^2}{\hbar^3}\frac{2p_1 \cdot p_2 + \mathcal{O}(\hbar)}{\wn q^2} \tilde{T}_1 \cdot \tilde{T}_2\,.\label{eqn:scalarYMamp}
\end{equation}
Upon substitution into the impulse in \eqn~\eqref{eqn:limImpulse} or angular impulse in \eqn~\eqref{eqn:limAngImp} the apparently singular denominator in the $\hbar\rightarrow 0$ limit is cancelled. It is only these quantities, not the amplitudes, that are classically well defined and observable.

\subsubsection*{Spinor-scalar}

We can include massive Dirac spinors $\psi$ in the Yang-Mills amplitudes by using a Lagrangian $\mathcal{L} = \mathcal{L}_0 + \mathcal{L}_{\textrm{Dirac}}$, where the Dirac Lagrangian 
\begin{equation}
\mathcal{L}_\textrm{Dirac} = \bar{\psi}\left(i\slashed{\partial} - m\right)\psi\label{eqn:DiracL}
\end{equation}
includes a minimal coupling to the gauge field. The tree level amplitude for spinor-scalar scattering is then
\begin{equation}
i\mathcal{A}^{ab}_{1/2-0} = \frac{i\tilde{g}^2}{2q^2}\bar{u}^a(p_1+q)\gamma^\mu u^{b}(p_1)(2p_2 - q)_\mu\, \tilde{T}_1\cdot \tilde{T}_2\,.
\end{equation}
We are interested in the pieces that survive to the classical limit. To extract them we must set the momentum transfer as $q = \hbar\wn q$ and expand the amplitude in powers of $\hbar$. 

The subtlety here is the on-shell Dirac spinor product. In the limit, when $q$ is small, we can follow the logic of \eqn~\eqref{eqn:infinitesimalBoost} and interpret $\bar{u}^a(p_1 + \hbar\wn q) \sim \bar{u}^a(p_1) + \Delta\bar{u}^a(p_1)$ as being infinitesimally Lorentz boosted, see also~\cite{Lorce:2017isp}. One expects amplitudes for spin $\nicefrac{1}{2}$ particles to only be able to probe up to linear order in spin (i.e. the dipole of a spinning body) \cite{Vaidya:2014kza,Guevara:2017csg,Guevara:2018wpp}, so in deriving the infinitesimal form of the Lorentz transformation we expand to just one power in the spin. The infinitesimal parameters $\omega_{\mu\nu}$ are exactly those determined in \eqn~\eqref{eqn:LorentzParameters}, so in all the leading terms of the spinor product are
\begin{equation}
\bar{u}^a(p_1 + \hbar\wn q) \gamma_\mu u^b(p_1) = 2{p_1}_\mu \delta^{ab} + \frac{\hbar}{4m^2}\bar{u}^a(p_1) p{_1}^\rho \wn q^\sigma [\gamma_\rho,\gamma_\sigma] \gamma_\mu u^b(p_1) + \mathcal{O}(\hbar^2)\,.\label{eqn:spinorshift}
\end{equation}
Evaluating the product of gamma matrices via the identity
\begin{equation}
[\gamma_\mu, \gamma_\nu] \gamma_\rho = 2\eta_{\nu\rho}\gamma_\mu - 2\eta_{\mu\rho}\gamma_\nu -  2i\epsilon_{\mu\nu\rho\sigma} \gamma^\sigma \gamma^5\,,\label{eqn:3gammaCommutator}
\end{equation}
where $\epsilon_{0123} = +1$ and $\gamma^5 = i\gamma^0 \gamma^1 \gamma^2 \gamma^3$, the spinor product is just
\begin{multline}
\bar{u}^a(p_1 + \hbar\wn q) \gamma_\mu u^b(p_1) = 2{p_1}_\mu\delta^{ab}  + \frac{\hbar}{2m_1^2} \bar{u}^a(p_1) p_1^\rho \wn q^\sigma \left(\gamma_\sigma\eta_{\mu\rho} - \gamma_\rho\eta_{\mu\sigma}\right) u^b(p_1) \\ - \frac{i\hbar}{2m_1^2} \bar{u}^a(p_1) p{_1}^\rho \wn q^\sigma \epsilon_{\rho\sigma\mu\delta} \gamma^\delta\gamma^5 u^b(p_1) + \mathcal{O}(\hbar^2)\,.
\end{multline}
Comparing with our result from \eqn~\eqref{eqn:spinorSpinVec}, the third term clearly hides an expression for the spin $\nicefrac{1}{2}$ polarisation vector. Making this replacement and substituting the spinor product into the amplitude yields, for on shell kinematics, only two terms at an order lower than $\mathcal{O}(\hbar^2)$:
\begin{equation}
\hbar^3\mathcal{A}^{ab}_{1/2-0} = \frac{2g^2}{\wn q^2} \left((p_1\cdot p_2)\delta^{ab} - \frac{i}{m_1^2} p{_1}^\rho \wn q^\sigma p_2^\lambda \epsilon_{\rho\sigma\lambda\delta} s^{\delta\,ab}_1 + \mathcal{O}(\hbar^2)\right) \tilde{T}_1\cdot\tilde{T}_2\,,\label{eqn:spinorYMamp}
\end{equation}
where we adopt the notation ${s_1}^{\mu}_{ab} = {s_1}^{\mu}_{ab}(p_1)$.

\subsubsection*{Vector-scalar}

Now consider scattering a massive vector rather than spinor. The minimally coupled gauge interaction can be obtained by applying the Higgs mechanism to the Yang-Mills Lagrangian, which when added to the Lagrangian $\mathcal{L}_0$ yields the tree-level amplitude
\begin{multline}
i\mathcal{A}^{ij}_{1-0} = -\frac{i\tilde{g}^2}{2q^2} \varepsilon{_i^*}^\mu(p_1+q) \varepsilon^\nu_j(p_1) \left(\eta_{\mu\nu}(2p_1+q)_\lambda - \eta_{\nu\lambda}(p_1-q)_\mu \right.\\\left. - \eta_{\lambda\mu}(2q+p_1)_\nu \right)(2p_2 - q)^\lambda\, \tilde{T}_1\cdot \tilde{T}_2\,.
\end{multline}
To obtain the classically significant pieces of this amplitude we must once more expand the product of on-shell tensors, in this case the polarisation vectors. In the classical limit we can again consider the outgoing polarisation vector as being infinitesimally boosted, so $\varepsilon{_i^*}^\mu(p_1 + \hbar\wn q) \sim \varepsilon{_i^*}^\mu(p_1) + \Delta\varepsilon{_i^*}^\mu(p_1)$. 

However, from spin 1 particles we expect to be able to probe $\mathcal{O}(s^2)$, or quadrupole, terms \cite{Vaidya:2014kza,Guevara:2017csg,Guevara:2018wpp}. Therefore it is salient to expand the Lorentz boost to two orders in the Lorentz parameters $\omega_{\mu\nu}$, so under infinitesimal transformations we take
\begin{equation}
\varepsilon_i^\mu(p) \mapsto \Lambda^\mu{ }_\nu\, \varepsilon_i^\nu(p) \simeq \left(\delta^\mu{ }_\nu - \frac{i}{2} \omega_{\rho\sigma} (\Sigma^{\rho\sigma})^\mu{ }_\nu - \frac18 \left((\omega_{\rho\sigma} \Sigma^{\rho\sigma})^2\right)^\mu{ }_\nu \right) \varepsilon_i^\nu(p)\,,
\end{equation}
where $(\Sigma^{\rho\sigma})^\mu{ }_\nu = i\left(\eta^{\rho\mu} \delta^\sigma{ }_\nu - \eta^{\sigma\mu} \delta^\rho{}_\nu\right)$. Since the kinematics are again identical to those used to derive \eqn~\eqref{eqn:LorentzParameters}, we get
\begin{equation}
\varepsilon{_i^*}^\mu(p_1 + \hbar\wn q)\, \varepsilon^\nu_j(p_1) = \varepsilon{_i^*}^\mu \varepsilon^\nu_j - \frac{\hbar}{m_1^2} (\wn q \cdot \varepsilon_i^*) p_1^\mu \varepsilon^\nu_j - \frac{\hbar^2}{2 m_1^2} (\wn q\cdot\varepsilon_i^*) \wn q^\mu \varepsilon^\nu_j + \mathcal{O}(\hbar^3)\,,\label{eqn:vectorS}
\end{equation}
where now $\varepsilon_i$ will always be a function of $p_1$, so in the classical limit $\varepsilon_i^*\cdot p_1 = \varepsilon_i\cdot p_1=0$. Using this expression in the full amplitude, the numerator becomes
\begin{multline}
n_{ij} = 2(p_1\cdot p_2)(\varepsilon_i^*\cdot\varepsilon_j) - 2\hbar (p_2\cdot\varepsilon_i^*)(\wn q\cdot\varepsilon_j) + 2\hbar (p_2\cdot\varepsilon_j)(\wn q\cdot\varepsilon_i^*) \\ + \frac{1}{m_1^2}\hbar^2 (p_1\cdot p_2)(\wn q\cdot\varepsilon_i^*)(\wn q\cdot\varepsilon_j) + \frac{\hbar^2}{2}\wn q^2 (\varepsilon_i^*\cdot \varepsilon_j) + \mathcal{O}(\hbar^3)\,.
\end{multline}
How the spin vector enters this expression is not immediately obvious, and relies on Levi-Civita tensor identities. At $\mathcal{O}(\hbar)$, $\epsilon^{\delta\rho\sigma\nu} \epsilon_{\delta\alpha\beta\gamma} = -3!\, \delta^{[\rho}{ }_\alpha \delta^{\sigma}{ }_\beta \delta^{\nu]}{ }_\gamma$ leads to
\begin{align}
\hbar (p_2\cdot\varepsilon{_i^*})(\wn q \cdot \varepsilon_j) - \hbar (p_2 \cdot \varepsilon_j)(\wn q \cdot \varepsilon_i^*) = \frac{\hbar}{m_1^2}&p_1^\rho \wn q^\sigma p_2^\lambda \epsilon_{\delta\rho\sigma\lambda}\epsilon^{\delta\alpha\beta\gamma}{\varepsilon_i^*}_\alpha {\varepsilon_j}_\beta {p_1}_\gamma \nonumber \\ &\equiv  -\frac{i}{m_1}p_1^\rho \wn q^\sigma p_2^\lambda \epsilon_{\rho\sigma\lambda\delta} {s_1}^\delta_{ij}\,,
\end{align}
where again we are able to identify the spin 1 polarisation vector calculated in \eqn~\eqref{eqn:vectorSpinVec} and introduce it into the amplitude. There is also a spin vector squared contribution entering at $\mathcal{O}(\hbar^2)$; observing this is reliant on applying the identity $\epsilon^{\mu\nu\rho\sigma} \epsilon_{\alpha\beta\gamma\delta} = -4!\, \delta^{[\mu}{ }_\alpha \delta^{\nu}{ }_\beta \delta^{\rho}{ }_\gamma \delta^{\sigma]}{ }_\delta$ and the expression in \eqn~\eqref{eqn:vectorSpinVec} to calculate
\begin{equation}
\sum_k \left(\wn q \cdot s_1^{ik}\right) (\wn q \cdot s_1^{kj}) = -\hbar^2 (\wn q \cdot \varepsilon_i^*) (\wn q\cdot \varepsilon_j) - \hbar^2 \wn q^2 \delta_{ij} + \mathcal{O}(\hbar^3)\,.
\end{equation}
This particular relationship is dependent on the sum over helicities $\sum_h {\varepsilon^*_h}^\mu \varepsilon^\nu_h = -\eta^{\mu\nu} + \frac{p_1^\mu p_1^\nu}{m_1^2}$ for massive vector bosons, an additional consequence of which is that $\varepsilon_i^*\cdot \varepsilon_j = -\delta_{ij}$. Incorporating these rewritings of the numerator in terms of spin vectors, the full amplitude is
\begin{multline}
\hbar^3\mathcal{A}^{ij}_{1-0} = \frac{2g^2}{\wn q^2}\left((p_1\cdot p_2)\delta^{ij} - \frac{i}{m_1} p{_1}^\rho \wn q^\sigma p{_2}^\lambda \epsilon_{\rho\sigma\lambda\delta} s_1^{\delta\,ij} + \frac{1}{2 m_1^2}(p_1\cdot p_2)(\wn q\cdot s_1^{ik}) (\wn q \cdot s_1^{kj}) \right.\\\left. - \frac{\hbar^2 \wn q^2}{4m_1^2}\left(2(p_1\cdot p_2) + m_1^2\right) + \mathcal{O}(\hbar^3)\right) \tilde{T}_1\cdot \tilde{T}_2\,.
\end{multline}
The internal sum over spin indices in the $\mathcal{O}(s^2)$ term will now always be left implicit. In classical observables we can also drop the remaining $\mathcal{O}(\hbar^2)$ term, as this just corresponds to a quantum correction from contact interactions \cite{Kosower:2018adc}.

\subsection{Gravity amplitudes}

Rather than recompute corresponding gravity amplitudes in perturbative GR\footnote{We have checked that direct calculations with graviton vertex rules given in \cite{Holstein:2008sx} reproduce our results.}, we can easily just apply the double copy. The generalisation of the traditional BCJ gauge theory replacement rules \cite{Bern:2008qj,Bern:2010ue} to massive matter states was developed by Johansson and Ochirov \cite{Johansson:2014zca}. In our context the colour-kinematics replacement is always trivial: all the amplitudes only have a $t$ channel diagram, and subsequently have identical colour factors. This makes the Jacobi identities trivial, so by just replacing colour factors with the desired numerator we are guaranteed to land on a gravity amplitude, provided we replace $g\rightarrow\frac{\kappa}{2}$, where $\kappa = \sqrt{32\pi G}$\footnote{Note that, analogously to \eqn~\eqref{eqn:scalarQCDLagrangian}, the coupling constant in the Einstein-Hilbert Lagrangian is $\tilde{\kappa} = \kappa/\sqrt{\hbar}$.}.
 
In particular, if we replace the colour factor in the previous spin $n$--spin 0 Yang-Mills amplitudes with the scalar numerator from \eqn~\eqref{eqn:scalarYMamp} we will obtain a spin $n$--spin 0 gravity amplitude, as the composition of little group irreps is simply $(\mathbf{2n + 1})\otimes\mathbf{1}=\mathbf{2n + 1}$. Using the scalar numerator ensures that the spin index structure passes to the gravity theory unchanged. Thus we can immediately obtain that the classically significant part of the spin $\nicefrac{1}{2}$--spin 0 gravity amplitude is\footnote{The overall sign is consistent with the replacements in \cite{Bern:2008qj,Bern:2010ue} for our amplitudes' conventions.}
\begin{equation}
\hbar^3 \mathcal{M}^{ab} = -\left(\frac{\kappa}{2}\right)^2\frac{4}{\wn q^2} \left[(p_1\cdot p_2)^2\delta^{ab} - \frac{i}{m_1}(p_1\cdot p_2) p_1^\rho \wn q^\sigma p_2^\lambda \epsilon_{\rho\sigma\lambda\alpha} s^{\alpha\,ab}_1 + \mathcal{O}(\hbar^2)\right],\label{eqn:spinorScalarGravAmp}
\end{equation}
while that for spin 1--spin 0 scattering is
\begin{multline}
\hbar^3\mathcal{M}^{ij} = -\left(\frac{\kappa}{2}\right)^2 \frac{4}{\wn q^2} \left[(p_1\cdot p_2)^2\delta^{ij} - \frac{i}{m_1}(p_1\cdot p_2) p_1^\rho \wn q^\sigma p_2^\lambda \epsilon_{\rho\sigma\lambda\delta} s^{\delta\,ij}_1 \right.\\\left. + \frac{1}{2 m_1^2}(p_1\cdot p_2)^2(\wn q\cdot s_1^{ik}) (\wn q \cdot s_1^{kj}) + \mathcal{O}(\hbar^2)\right]. \label{eqn:vectorScalarGravAmp}
\end{multline}
Notice that the $\mathcal{O}(s)$ parts of these amplitudes are exactly equal, up to the different spin indices. This is a manifestation of gravitational universality: the gravitational coupling to the spin dipole should be independent of the spin of the field, precisely as we observe.

We have deliberately not labelled these as Einstein gravity amplitudes, because the gravitational modes in our amplitudes contain both gravitons $h_{\mu\nu}$ and scalar dilatons $\phi$. To see this, examine the factorisation channels in the $t$ channel cut of the vector amplitude:
\begin{multline}
\lim\limits_{\wn q^2 \rightarrow 0} \left(\wn q^2 \hbar^3 \mathcal{M}^{ij}\right) = -4\left(\frac{\kappa}{2}\right)^2\, \left(p_1^\mu p_1^{\tilde{\mu}} \delta^{ij} - \frac{i}{m_1}p_1^\mu \epsilon^{\tilde{\mu}\rho\sigma\delta} {p_1}_\rho \wn q_\sigma {s_1}_\delta^{ij} + \frac{1}{2m_1^2} (\wn q\cdot s_1^{ik})(\wn q\cdot s_1^{kj}) p_1^\mu p_1^{\tilde{\mu}}\right) \\ \times \mathcal{P}^{(4)}_{\mu\tilde{\mu}\nu\tilde{\nu}}\, p_2^\nu p_2^{\tilde{\nu}} - 4\left(\frac{\kappa}{2}\right)^2\left(p_1^\mu p_1^{\tilde{\mu}} \delta^{ij} + \frac{(\wn q\cdot s_1^{ik})(\wn q\cdot s_1^{kj})}{2 m_1^2} p_1^\mu p_1^{\tilde{\mu}}\right) \mathcal{D}^{(4)}_{\mu\tilde{\mu}\nu\tilde{\nu}}\, p_2^\nu p_2^{\tilde{\nu}}\,,
\end{multline}
where
\begin{equation}
\mathcal{P}^{(d)}_{\mu\tilde{\mu}\nu\tilde{\nu}} = \eta_{\mu(\nu}\eta_{\tilde{\nu})\tilde{\mu}} - \frac{1}{d-2}\eta_{\mu\tilde{\mu}}\eta_{\nu\tilde{\nu}} \qquad \text{and} \qquad
\mathcal{D}^{(d)}_{\mu\tilde{\mu}\nu\tilde{\nu}} = \frac{1}{d-2}\eta_{\mu\tilde{\mu}}\eta_{\nu\tilde{\nu}}
\end{equation}
are the $d$-dimensional de-Donder gauge graviton and dilaton projectors respectively. The pure Einstein gravity amplitude for classical spin 1--spin 0 scattering can now just be read off as the part of the amplitude contracted with the graviton projector. We find that
\begin{multline}
\hbar^3\mathcal{M}^{ij}_{1-0} = -\left(\frac{\kappa}{2}\right)^2 \frac{4}{\wn q^2} \left[\left((p_1\cdot p_2)^2 - \frac12 m_1^2 m_2^2\right)\delta^{ij} - \frac{i}{m_1}(p_1\cdot p_2) p_1^\rho \wn q^\sigma p_2^\lambda \epsilon_{\rho\sigma\lambda\delta} s_1^{\delta\,ij} \right.\\\left. + \frac{1}{2 m_1^2}\left((p_1\cdot p_2)^2 - \frac12 m_1^2 m_2^2\right)(\wn q\cdot s_1^{ik}) (\wn q \cdot s_1^{kj}) + \mathcal{O}(\hbar^2)\right]. \label{eqn:vectorGravAmp}
\end{multline}
The spinor-scalar Einstein gravity amplitude receives the same correction to the initial, scalar component of the amplitude. 

Note that dilaton modes are coupling to the scalar monopole and $\mathcal{O}(s^2)$ quadrapole terms in the gravity amplitudes, but not to the $\mathcal{O}(s)$ dipole component. We also do not find axion modes, as observed in previous applications of the classical double copy to spinning particles \cite{Li:2018qap,Goldberger:2017ogt}, because axions are unable to couple to the massive external scalar.

\section{Black hole scattering observables from amplitudes}
\label{sec:KerrCalcs}

We are now armed with a set of classical tree-level amplitudes and formulae for calculating the linear impulse $\Delta p_1^\mu$ and angular impulse $\Delta s_1^\mu$ from them. We also already have a clear target where the analogous classical results are known: the results for 1PM scattering of spinning black holes found in \cite{Vines:2017hyw}. 

Given our amplitudes only reach the quadrupole level, we can only probe lower order terms in the expansion of \eqn~\eqref{eqn:KerrDeflections}. Expanding in the rescaled spin $a_1^\mu$, and setting $a_2^\mu\to0$, the linear impulse is
\begin{multline}
\Delta p_1^{\mu} = \frac{2 G m_1m_2}{\sqrt{\gamma^2 - 1}} \left\{(2\gamma^2 - 1) \frac{{b}^\mu}{b^2} + \frac{2\gamma}{b^4} \Big( 2{b}^\mu {b}^\nu-b^2\Pi^{\mu\nu}\Big) \epsilon_{\nu\rho\alpha\beta} u_1^\alpha u_2^\beta a_1^\rho \right.
\\
\left. - \frac{2\gamma^2 - 1}{{b}^6} \Big(4b^\mu b^\nu b^\rho-3b^2 b^{(\mu}\Pi^{\nu\rho)}\Big)a_{1\nu}a_{1\rho} + \mathcal{O}(a^3) \right\}+\mathcal O(G^2), \label{eqn:JustinImpResult}
\end{multline}
where $\Pi^\mu{}_\nu$ is the projector into the plane orthogonal to $u_1^\mu$ and $u_2^\mu$ from \eqref{eqn:projector}.
Meanwhile the angular impulse to the same order is
\begin{equation}
\begin{gathered}
\Delta s_1^{\mu} =-u_1^\mu a_{1\nu}\Delta p_1^\nu -\frac{2G m_1m_2}{\sqrt{\gamma^2-1}}\left\{
 2\gamma \epsilon^{\mu\nu\rho\sigma}  u_{1\rho} \epsilon_{\sigma\alpha\beta\gamma} u_1^\beta u_2^\gamma \frac{{b}^\alpha}{b^2}a_{1\nu}  \right.
\\
\left. - \frac{2\gamma^2 - 1}{b^4} \epsilon^{\mu\nu\kappa\lambda} u_{1\kappa}  \Big( 2{b}_\nu {b}_\rho-b^2\Pi_{\nu\rho} \Big)a_{1\lambda} a_1^\rho 
 + \mathcal{O}(a^3) \right\}+\mathcal O(G^2). \label{eqn:JustinSpinResult}
\end{gathered}
\end{equation}
In this section we demonstrate that both of these results can be recovered by using the classical pieces of our Einstein-gravity amplitudes.

\subsection{Linear impulse}

To calculate the linear impulse we substitute $\mathcal{M}_{1-0}$ into the general expression in \eqn~\eqref{eqn:limImpulse}. Following the prescription in \sect{sec:classicalLim}, the only effect of the momentum integrals in the expectation value is to set $p_i \rightarrow m_iu_i$ in the classical limit. This then reduces the double angle bracket to the single expectation value over the spin states:
\begin{equation}
\begin{aligned}
\Delta p_1^{\mu,(0)} &= -i m_1 m_2 \left(\frac{\kappa}{2}\right)^2 \int\!\dd^4\wn q\, \del(u_1\cdot\wn q) \del(u_2\cdot\wn q) e^{-ib\cdot\wn q} \frac{\wn q^\mu}{\wn q^2} \\&\qquad\times \left\langle \frac12(2\gamma^2 - 1) - i \gamma u_1^\rho \wn q^\sigma u_2^\nu \epsilon_{\rho\sigma\nu\delta}\, \frac{s_1^\delta}{m_1} + \frac{2\gamma^2 - 1}{4m_1^2} (\wn q\cdot s_1) (\wn q \cdot s_1) \right\rangle\\
&\equiv -4i m_1 m_2 \pi G \left((2\gamma^2 - 1) I^\mu  -  2i\gamma u_1^\rho u_2^\nu \epsilon_{\rho\sigma\nu\delta}\, \big\langle a_1^\delta \big\rangle I^{\mu\sigma} + \frac{2\gamma^2 - 1}{2} \big\langle {a_1}_\nu {a_1}_\rho \big\rangle I^{\mu\nu\rho} \right),
\end{aligned}
\end{equation}
where we have rescaled $a^\mu = s^\mu/m$ and defined three integrals of the general form
\begin{equation}
I^{\mu_1\cdots \mu_n} = \int\!\dd^4\wn q\, \del(u_1\cdot\wn q) \del(u_2\cdot \wn q) \frac{e^{-ib\cdot\wn q}}{\wn q^2} \wn q^{\mu_1} \cdots \wn q^{\mu_n}\,.\label{eqn:defOfI}
\end{equation}

The lowest rank integral of this type was evaluated in \cite{Kosower:2018adc}, with the result
\begin{equation}
I^\mu = \frac{i}{2\pi \sqrt{\gamma^2 - 1}} \frac{{b}^\mu}{b^2}\,,\label{eqn:I1result}
\end{equation}
To evaluate the higher rank examples, note that the results must lie in the plane orthogonal to the four velocities. This plane is spanned by the impact parameter $b^\mu$, and the projector $\Pi^\mu{ }_\nu$ defined in \eqn~\eqref{eqn:projector}. Thus, for example,
\begin{equation}
I^{\mu\nu} = \alpha_2 b^\mu b^\nu + \beta_2 \Pi^{\mu\nu}\,.
\end{equation}
Given that we are working away from the threshold value $b = 0$, the left hand side is traceless and $\beta_2 = - \alpha_2\, b^2 /2$. Then contracting both sides with $b_\nu$, one finds
\begin{equation}
\alpha_2 b^2\, b^\mu = 2\int\!\dd^4\wn q\, \del(u_1\cdot\wn q) \del(u_2\cdot \wn q) \frac{e^{-ib\cdot\wn q}}{\wn q^2} \wn q^{\mu} (b\cdot\wn q) = \frac{1}{\pi \sqrt{\gamma^2 - 1}} \frac{b^\mu}{b^2}\,,
\end{equation} 
where we have used the result of \eqn~\eqref{eqn:I1result}. Thus the coefficient $\alpha_2$ is uniquely specified, and we find
\begin{equation}
I^{\mu\nu} = \frac{1}{\pi b^4 \sqrt{\gamma^2 - 1}} \left(b^\mu b^\nu - \frac12 b^2 \Pi^{\mu\nu} \right)\label{eqn:I2result}.
\end{equation}
Following an identical procedure for $I^{\mu\nu\rho}$, we can then readily determine that
\begin{equation}
I^{\mu\nu\rho} = -\frac{4i}{\pi b^6 \sqrt{\gamma^2 - 1} } \left(b^\mu b^\nu b^\rho - \frac34 b^2 b^{(\mu} \Pi^{\nu\rho)} \right).\label{eqn:I3result}
\end{equation}

Substituting the integral results into the expression for the leading order classical impulse, and expanding the projectors from \eqn~\eqref{eqn:projector}, then leads to
\begin{multline}
\Delta p_1^{\mu,(0)} = \frac{2G m_1 m_2}{\sqrt{\gamma^2 - 1}}\left((2\gamma^2 - 1) \frac{{b}^\mu}{b^2} + \frac{2\gamma}{b^4} ( 2 b^\mu {b}^\alpha-b^2\Pi^{\mu\alpha}) \epsilon_{\alpha\rho\sigma\delta} u_1^\rho u_2^\sigma \big\langle a_1^\delta \big\rangle  \right.
\\
\left.- \frac{2\gamma^2 - 1}{{b}^6} (4b^\mu b^\nu b^\rho-3b^2 b^{(\mu}\Pi^{\nu\rho)})\langle a_{1\nu}a_{1\rho}\rangle\right).\label{eqn:QuantumImpRes}
\end{multline}
Comparing with \eqn~\eqref{eqn:JustinImpResult} we observe an exact match, up to the appearance of spin state expectation values, between our result and the $\mathcal{O}(a^2)$ expansion of the result for spinning black holes from \cite{Vines:2017hyw}. 

\subsection{Angular impulse}

Our expression, equation~\eqref{eqn:limAngImp}, for the classical leading-order angular impulse naturally has two parts: one term has a commutator while the other term does not. For clarity we will handle these two parts separately, beginning with the term without a commutator---which we will call the direct term.

\subsubsection*{The direct term}

Substituting our $\mathcal{O}(s^2)$ Einstein-gravity amplitude, equation~\eqref{eqn:vectorGravAmp}, into the direct part of the general angular impulse formula, we find
\[
\Delta s_1^{\mu,(0)}\big|_{\textrm{direct}} 
&\equiv
\Lexp i\! \int\!\dd^4\wn q\,\del(2p_1\cdot\wn q) \del(2p_2\cdot\wn q)e^{-ib\cdot\wn q} \bigg(-\hbar^3 \frac{p_1^\mu}{m_1^2} \qb \cdot \s_{1} (p_1) \mathcal{M}_{1-0} \bigg) \Rexp \\
= \Lexp \frac{i \kappa^2}{m_1^2} &\int\! \dd^4\wn q\,\del(2p_1\cdot\wn q) \del(2p_2\cdot\wn q) \frac{e^{-ib\cdot\wn q}}{\qb^2}\, p_1^\mu \wn q \cdot \s_1(p) \bigg(\bigg((p_1\cdot p_2)^2  - \frac{1}{2}m_1^2m_2^2\bigg) \\ &\qquad\qquad\qquad\qquad - \frac{i}{m_1}(p_1\cdot p_2)p_1^\alpha \wn q^\beta p_2^\gamma \epsilon_{\alpha\beta\gamma\delta} \, s_1^\delta(p) \bigg)+ \mathcal{O}(s^3) \Rexp\,.
\]
As with the linear impulse, we can reduce the double angle brackets to single, spin state, angle brackets by replacing $p_i \rightarrow m_i u_i$, so that
\begin{equation}
\Delta s^{\mu,(0)}_1\big|_{\textrm{direct}} = 4\pi G m_2\, u_1^\mu  \left( i  \left(2\gamma^2 - 1\right) \langle s_1^\nu \rangle I_\nu + \frac{2}{m_1} \gamma 
\, u_1^\alpha u_2^\gamma \epsilon_{\alpha\beta\gamma\delta} \langle s_{1\nu}  s_1^{\delta} \rangle \, I^{\nu\beta} \right)\,,
\end{equation}
where the integrals are again defined by \eqn~\eqref{eqn:defOfI}. We can now just substitute our previous evaluations of these integrals, equations~\eqref{eqn:I1result} and~\eqref{eqn:I2result}, to learn that
\begin{equation}
\Delta a_1^{\mu,(0)}\big|_{\textrm{direct}} = -\frac{2G m_2}{\sqrt{\gamma^2 -1}} u_1^\mu \left((2\gamma^2 - 1)\frac{{b}_\nu}{b^2} \spinExp{a_1^\nu} + \frac{2\gamma}{b^4} \left( 2{b}^\nu {b}^\alpha-{b^2}\Pi^{\nu\alpha}\right) \epsilon_{\alpha\beta\gamma\delta} u_1^\beta u_2^\gamma \spinExp{{a_1}_\nu a_1^\delta} \right).\label{eqn:linPiece}
\end{equation}

\subsubsection*{The commutator term}

Now we turn to the commutator piece of \eqn~\eqref{eqn:limAngImp}. The scalar part of our Einstein-gravity amplitude, equation~\eqref{eqn:vectorGravAmp}, has diagonal spin indices, so its commutator vanishes. We encounter two non-vanishing commutators:
\[
[\s_1^\mu, \s_1^\delta]&=-i \hbar \, \epsilon^{\mu\delta\rho\sigma} s_{1\,\rho} \frac{p_{1 \, \sigma}}{m_1} \,,\\
[\s_1^\mu, \qb \cdot \s_1 \, \qb\cdot \s_1] &= -2 i \hbar \, \qb \cdot s_1 \, \epsilon^{\mu\alpha\beta\gamma} \qb_\alpha s_{1\, \beta} \, \frac{p_{1\, \gamma}}{m_1} + \mathcal{O}(\hbar^2) \,,
\]
omitting a term which is higher order. Using these expressions in the commutator term, the result is
\[
\hspace{-10pt}
\Delta s_1^{\mu, (0)}|_\textrm{com} &= i\, \Lexp \int \dd^4 \qb \, \del(2p_1 \cdot \qb) \del(2p_2 \cdot \qb) e^{-ib \cdot \qb} \, \hbar^2 [s^\mu(p), \mathcal{M}_{1-0}] \Rexp \\
&= i \kappa^2 \Lexp \int \dd^4 \qb \, \del(2p_1 \cdot \qb) \del(2p_2 \cdot \qb) \frac{e^{-ib \cdot \qb}}{\qb^2} \bigg( (p_1 \cdot p_2) p_1^\alpha \qb^\beta p_2^\gamma \epsilon_{\alpha\beta\gamma\sigma} \epsilon^{\mu\nu\rho\sigma} s_{1 \nu} \frac{p_{1 \rho}}{m_1^2}  \\
& \qquad\qquad\qquad\qquad+ \frac{i}{m_1^3} \left( (p_1\cdot p_2)^2 - \frac12 m_1^2 m_2^2\right) \qb \cdot \s_1 \, \epsilon^{\mu\nu\rho\sigma} \qb_\nu s_{1\sigma}p_{1\rho} \bigg) \Rexp.
\]
As is familiar by now, we evaluate the integrals over the momentum-space wave functions by setting $p_i  = m_i u_i$, but expectation values over the spin-space wave functions remain. The result can be organised in terms of the integrals $I^\alpha$ and $I^{\alpha\beta}$ defined in equation~\eqref{eqn:defOfI}:
\[
\Delta s_1^{\mu, (0)}|_\textrm{com} &= 2\pi i \, G m_2 \bigg( 4\gamma \epsilon^{\mu\nu\rho\sigma} \langle s_{1\,\nu} \rangle u_{1\,\rho} \epsilon_{\sigma\alpha\beta\gamma} u_1^\beta u_2^\gamma I^\alpha 
- \frac{2 i}{m_1} (2\gamma^2 - 1) \epsilon^{\mu\nu\rho\sigma} u_{1\, \rho} \langle s_{1\, \sigma} {s_{1}}^\alpha \rangle I_{\alpha\nu}
\bigg) \, .
\]
Finally, we perform the integrals using equations~\eqref{eqn:I1result} and~\eqref{eqn:I2result}, rescale the spin vector to $a_1^\mu$ and combine the result with the direct contribution in \eqn~\eqref{eqn:linPiece}, to find that the angular impulse at $\mathcal{O}(a^2)$ is
\begin{multline}
\Delta s_1^{\mu,(0)}
= -\frac{2Gm_1 m_2}{\sqrt{\gamma^2 -1}} \bigg\{(2\gamma^2 - 1) u_1^\mu \frac{{b}_\nu}{b^2} \spinExp{a_1^\nu} - \frac{2\gamma}{b^2} u_1^\mu \left(\eta^{\nu\alpha} - \frac{2{b}^\nu {b}^\alpha}{b^2}\right) \epsilon_{\alpha\beta\gamma\delta} u_1^\beta u_2^\gamma \spinExp{{a_1}_\nu a_1^\delta} \\  + \frac{(2\gamma^2 - 1)}{b^2} \epsilon^{\mu\nu\rho\sigma} {u_1}_\rho \spinExp{{a_1}_\sigma {a_1}_\lambda} \left(\Pi^\lambda{}_\nu - \frac{2{b}_\nu {b}^\lambda}{b^2}\right)  
+ 2\gamma \epsilon^{\mu\nu\rho\sigma} \spinExp{a_{1\,\nu}} u_{1\,\rho} \epsilon_{\sigma\alpha\beta\gamma} u_1^\beta u_2^\gamma \frac{{b}^\alpha}{b^2} \bigg\}\,.
\label{eqn:QuantumAngImpRes}
\end{multline}
This final result agrees in detail with the classical result of equation~\eqref{eqn:JustinSpinResult}, modulo the remaining spin expectation values.

\section{Discussion}
\label{sec:discussion}

Starting from a quantum field theory for massive spinning particles with arbitrary long-range interactions (mediated e.g.\ by gauge bosons or gravitons), we have followed a careful analysis of the classical limit $(\hbar\to0)$ for long-range scattering of spatially localized wavepackets.   We have thereby arrived at fully relativistic expressions for the linear and angular impulses, the net changes in the linear and intrinsic angular momenta of the massive particles, due to an elastic two-body scattering process.  These, our central results, expressed in terms of on-shell scattering amplitudes, are given explicitly at leading order in the coupling by \eqref{eqn:limImpulse} and \eqref{eqn:limAngImp}.  Our general formalism places no restrictions on the order in coupling, and the expression \eqref{eqn:spinShift} for the angular impulse, like its analog for the linear impulse found in \cite{Kosower:2018adc}, should hold at all orders.

We have applied these general results to the examples of a massive spin 1/2 or spin 1 particle (particle 1) exchanging gravitons with a massive spin 0 particle (particle 2), imposing minimal coupling.  The results for the linear and angular impulses for particle 1, $\Delta p_1^\mu$ and $\Delta s_1^\mu$, due to its scattering with the scalar particle 2, are given by \eqref{eqn:QuantumImpRes} and \eqref{eqn:QuantumAngImpRes}.  These expressions are valid to linear order in the gravitational constant $G$, or to 1PM order, having arisen from the tree-level on-shell amplitude for the two-body scattering process.  By momentum conservation (in absence of radiative effects at this order), $\Delta p_2^\mu=-\Delta p_1^\mu$, and the scalar particle has no intrinsic angular momentum, $s_2^\mu=\Delta s_2^\mu=0$.  The spin 1/2 case provides the terms through linear order in the rescaled spin $a_1^\mu=s_1^\mu/m_1$, and the spin 1 case yields the same terms through linear order plus terms quadratic in $a_1^\mu$.

Our final results \eqref{eqn:QuantumImpRes} and \eqref{eqn:QuantumAngImpRes} from the quantum analysis are seen to be in precise agreement with the results \eqref{eqn:JustinImpResult} and \eqref{eqn:JustinSpinResult} from \cite{Vines:2017hyw} for the classical scattering of a spinning black hole with a nonspinning black hole, through quadratic order in the spin---except for the appearance of spin-state expectation values $\langle a_1^\mu\rangle$ and $\langle a_1^\mu a_1^\nu\rangle$ in the quantum results replacing $a_1^\mu$ and $a_1^\mu a_1^\nu$ in the classical result.  For any quantum states of a finite-spin particle, these expectation values cannot satisfy the  appropriate properties of their classical counterparts, e.g., $\langle a^\mu a^\nu\rangle \ne \langle a^\mu\rangle \langle a^\nu\rangle$.  Furthermore, we know that the intrinsic angular momentum of a quantum spin $n$ particle scales like $\langle s^\mu\rangle=m\langle a^\mu\rangle\sim n\hbar$, and we would thus actually expect any spin effects to vanish in a classical limit where we take $\hbar\to0$ at fixed spin quantum number $n$.  A fully consistent classical limit yielding nonzero contributions from intrinsic spin would need to take $n\to\infty$ as $\hbar\to0$, to keep $\langle s^\mu\rangle\sim n\hbar$ finite.  

However, the expansions in spin operators of the minimally coupled amplitudes and impulses, expressed in the forms we have derived here, are found to be universal, in the sense that going to higher spin quantum numbers $n$ continues to reproduce the same expressions at lower orders in the spin operators.  We have seen this explicitly here for the linear-in-spin level, up to spin 1, and the results of \cite{Guevara:2018wpp,Bautista:2019tdr,Chung:2018kqs} strongly suggest that an application of our formalism to minimally coupled amplitudes for arbitrary spin $n$ will confirm this pattern.  Furthermore, as $n\to\infty$, the spin states can indeed approach the limit where $\langle a^\mu a^\nu\rangle = \langle a^\mu\rangle \langle a^\nu\rangle$ and so forth.  We leave an analysis of higher spins for future work.

Our formalism provides a direct link between gauge-invariant quantities, on-shell amplitudes and classical asymptotic scattering observables, with generic incoming and outgoing states for relativistic spinning particles.  It is tailored to be combined with powerful modern techniques for computing relevant amplitudes, such as unitarity methods as the double copy.  Already with our examples at the spin 1/2 and spin 1 levels, we have seen that it produces new evidence (for generic spin orientations, and without taking the nonrelativistic limit) for the beautiful correspondence between classical spinning black holes and massive spinning quantum particles which are minimally coupled to gravity, first noted in \cite{Vaidya:2014kza}.  We look forward to future investigations of the extent to which this correspondence holds at higher orders, and to the possibility of its use in producing new results relevant to the dynamics of astrophysical binary black holes.

 
\section*{Acknowledgements}

We are grateful to Richard Ball, Lucile Cangemi, John-Joseph Carrasco, Thibault Damour, Alfredo Guevara, David Kosower and Alexander Ochirov for helpful discussions. BM and DOC thank IHES for hospitality during the completion of this work. BM is supported by STFC studentship ST/R504737/1. DOC is an IPPP associate, and thanks the IPPP for on-going support. He is supported in part by the STFC consolidated grant ``Particle Physics at the Higgs Centre''.

\appendix

\section{Conventions}

We adopt a mostly minus metric signature, with $\epsilon_{0123} = +1$. We absorb factors of $2\pi$ in integrand measures by letting $\dd^n x = d^n x/(2\pi)^n$, and in delta functions by $\del(x) = 2\pi \delta(x)$. Following \cite{Kosower:2018adc}, our conventions for Fourier transforms are then
\begin{equation}
f(x) = 	\int\! \dd^4\wn q\, \tilde{f}(\wn q) e^{-i\wn q\cdot x}\,, \qquad \tilde{f}(\wn q) = \int\! d^4x\, f(x) e^{i\wn q\cdot x}\,.
\end{equation}
We also adopt the convention that the Lorentz invariant phase space measure
\begin{equation}
\df(k) \equiv \dd^4 k\, \del^{(+)} (k^2 - m^2)\,,
\end{equation}
and that $\del_\Phi(k) \equiv 2E_k\, \del(\v{k})$, where $\v{k}$ is the spatial 3-vector defining the spatial components of the 4-vector $k^\mu$.

For a given tensor $X$, total symmmetrisation and antisymmetrisation respectively of tensor indices are represented by
\begin{equation}
\begin{aligned}
X^{(\mu_1} \dots X^{\mu_n)} &= \frac1{n!}\left(X^{\mu_1} X^{\mu_2} \dots X^{\mu_n} + X^{\mu_2} X^{\mu_1} \dots X^{\mu_n} + \cdots\right)\\
X^{[\mu_1} \dots X^{\mu_n]} &= \frac1{n!}\left(X^{\mu_1} X^{\mu_2} \dots X^{\mu_n} - X^{\mu_2} X^{\mu_1} \dots X^{\mu_n} + \cdots\right).
\end{aligned}
\end{equation}

Our definition of the amplitude differs by a phase factor relative to the standard definition used for the double copy. Here, in either gauge theory or gravity
\begin{equation}
i\mathcal{A}(p_1, p_2 \rightarrow p_1 + q, p_2 - q) = \sum \left(\text{Feynman diagrams}\right)\,,
\end{equation}
whereas in the convention used in \cite{Bern:2008qj,Bern:2010ue} the entire left hand side is defined as the amplitude. This means that one must incorporate extra factors of $i$ in our BCJ numerators, which leads to an overall minus sign upon squaring in the double copy.

\section{Explicit evaluation of the QFT spin vector}
\label{sec:spinVecDetails}

We have argued that the spin vector naturally emerges in quantum field theory as the expectation value of the Pauli-Lubanski operator. On physical particle states the inherent representation dependence is isolated by the spin polarisation vector $s_{ij}^\mu$, which was defined in \eqn~\eqref{eqn:PLinnerProd}. Here we develop the tools required to explicitly evaluate this equation, and explicitly derive the results for spin 1/2 and spin 1 particles.

The definition of the Pauli-Lubanski operator in \eqn~\eqref{eqn:PLvec} depends on the translation and Lorentz generators. In the quantum theory, the Noether charges associated with their respective symmetries can be used to construct explicit field operators. For example, translation symmetry of the Lagrangian $\mathcal{L}_\Psi$ leads to the existence of a conserved current, the canonical energy momentum tensor $\Theta^{\mu\nu}$. We can then represent the translation generator by the Noether charge
\begin{equation}
\mathbb{P}^\mu = \int\! d^3x\, \Theta^{0\mu} = \int\! d^3x \left(\Pi_s \partial^\mu \Psi_s - \eta^{0\mu}\mathcal{L}_{\Psi}\right)\label{eqn:momNoetherCharge}\,.
\end{equation}
Here $\Pi_s = \partial{L}_\Psi/\partial \dot{\Psi}_s$ is the canonical momentum, with $\dot{\Psi}_s \equiv \partial_0 \Psi_0$. The eigenvalues of the (normal ordered) field operator promotion of $\mathbb{P^\mu}$ then define the momentum of a particle.

In the same manner, angular momentum emerges from the conserved charge associated with Lorentz symmetry. The conserved Noether current is now \cite{Bogoliubov:1980,Weinberg:1995mt}
\begin{equation}
M^{\alpha\mu\nu} = x^\mu T^{\alpha\nu} - x^\nu T^{\alpha\mu}
\end{equation}
where $T^{\mu\nu}$ is the Belinfante tensor, the manifestly symmetric generalisation of the canonical $\Theta^{\mu\nu}$ which sources the gravitational field\footnote{In the sense that the Belifante tensor can be obtained by variations of the action with respect to the metric. It is given by $T^{\mu\nu} = \Theta^{\mu\nu} + \frac{i}{2}\partial_\alpha \left(\frac{\partial \mathcal{L}_{\Psi}}{\partial(\partial_\alpha \Psi_s)} \mathcal{S}^{\mu\nu} \Psi_s - \frac{\partial \mathcal{L}_{\Psi}}{\partial(\partial_\mu \Psi_s)} \mathcal{S}^{\alpha\nu} \Psi_s - \frac{\partial \mathcal{L}_{\Psi}}{\partial(\partial_\nu \Psi_s)} \mathcal{S}^{\alpha\mu} \Psi_s\right)$}. The associated charges $\int\! d^3x M^{0\mu\nu}$ take the form
\begin{equation}
\mathbb{J}^{\mu\nu} = \int\! d^3x \left( x^\mu \Theta^{0\nu} - x^\nu \Theta^{0\mu} + i \Pi_s \mathcal{S}^{\mu\nu} \Psi_s \right)
\equiv \mathbb{L}^{\mu\nu} + \mathbb{S}^{\mu\nu}.\label{eqn:angmomNoetherCharge}
\end{equation}
Eigenvalues of the operator promotion of this charge then define the angular momenta of a particle. The two terms correspond to orbital and intrinsic angular momenta, but as in GR we cannot uniquely make this splitting; only the total angular momentum is a well defined, conserved charge.

To uniquely obtain information about the pure spin part of $\mathbb{J}^{\mu\nu}$ we need to isolate the second term. This job is performed automatically by the Pauli-Lubanski operator: orbital contributions always drop out in its expectation values. These observables must then define the physical quantity which holds complete information about the intrinsic spin: the spin vector, $s^\mu$.

To see that this holds for any causal field of spin $s$, we can use the Fourier expansions of the field operators  \cite{Weinberg:1995mt},
\begin{equation}
\begin{aligned}
{\Psi}_s(x) =& \sum_\alpha \int\! \df(k) \left(a_\alpha(k) U_\alpha(k) e^{-ik\cdot x} + b^\dagger_\alpha(k) V_\alpha(k) e^{ik\cdot x}\right)\\
{\Pi}_s(x) =& \sum_\alpha \int\! \df(k) \left(b_\alpha(k) Y_\alpha(k) e^{-ik\cdot x} + a^\dagger_\alpha(k) X_\alpha(k) e^{ik\cdot x}\right).
\end{aligned}\label{eqn:FourierModes}
\end{equation}
Here $\alpha$ is the little group index, $a^\dagger_\alpha(k)$ and $b^\dagger_\alpha(k)$ are particle and antiparticle creation operators acting on the associated Fock space, and the momentum space tensors are in the same Lorentz representation as the field. Note that the canonical momentum operator's tensors are, by definition, dependent on those in ${\Psi}_s$. 

We can now expand the angular momentum operator $\mathbb{J}^{\mu\nu}$. We know from \eqn~\eqref{eqn:momNoetherCharge} that $\mathbb{L}^{\mu\nu}$ contains spatial derivatives - these will act on the Fourier modes in \eqn~\eqref{eqn:FourierModes}, so are replaced by 4-momenta. Thus the inner product
\begin{multline}
\langle p', a|\mathbb{J}_{\mu\nu}|p, b\rangle = i\langle 0| a_a(p') :\! \sum_{\alpha,\beta}\! \int\! \frac{\df(k)}{2E_k} \left(X_\alpha(k) \mathcal{S}_{\mu\nu} U_\beta(k)\, a^\dagger_\alpha(k)\, a_\beta(k) \right.\\\left. - 2x_{[\mu} k_{\nu]} X_\alpha(k) U_\beta(k)\, a^\dagger_\alpha(k)\, a_\beta(k) + \cdots\right)\! : a_b^\dagger(p) |0\rangle \,,\label{eqn:angMomOpInnerProd}
\end{multline}
where the ellipsis denotes terms containing operators $b(p)$. By virtue of the (anti) commutation relations all such terms do not contribute, since the antiparticle Fock space operators always annihilate the vacuum. The terms in \eqn~\eqref{eqn:momNoetherCharge} with explicit appearances of the Lagrangian have also disappeared; because the Lagrangian can always be written in terms of the field equations, it vanishes on physical states.

Let us restrict our attention to the orbital term on the second line, which corresponds to the inner product of $\mathbb{L}_{\mu\nu}$. How does this contribute to inner products of the Pauli-Lubanski operator, such as \eqn~\eqref{eqn:defOfQFTspinVec}? Since on momentum eigenstates $\mathbb{P}_\mu |p, s\rangle = p_\mu |p, s\rangle$, we will have
\begin{multline}
\epsilon^{\mu\nu\rho\sigma} \langle p', a|\mathbb{P}_\nu\mathbb{L}_{\rho\sigma}|p, b\rangle = i\sum_{\alpha, \beta} \!\int\! \frac{\df(k)}{E_k} \epsilon^{\mu\nu\rho\sigma} p'_\nu k_\rho x_\sigma X_\alpha(k) U_\beta(k)\, \langle 0|a_a(p') a^\dagger_\alpha(k) a_\beta(k) a_b^\dagger(p) |0\rangle\\ = \frac{i}{E_p'} \epsilon^{\mu\nu\rho\sigma} p'_\nu p'_\rho x_\sigma X_a(p') U_b(p') \del(p - p') = 0.
\end{multline}

Expectation values of $\mathbb{W}^\mu$ therefore receive contributions only from the intrinsic spin part of $\mathbb{J}^{\rho\sigma}$.  In particular, the only terms emerging from the vacuum expectation value in \eqn~\eqref{eqn:angMomOpInnerProd} equal $\del_\Phi(p' - k) \delta_{a\alpha} \del_\Phi(k - p) \delta_{\beta b}$; evaluating the phase space integral, this is just $\del_\Phi(p - p')$. Since the only 4-vector encapsulating the information about a particle's spin is the spin vector $s^\mu$, we must have that
\[
\langle p', j| \W^\mu |p, i \rangle \equiv m \s^\mu_{ij}(p)\, \del_\Phi(p-p') \,,
\]
which is exactly our definition in \eqn~\eqref{eqn:PLinnerProd}.

We now have all the tools needed to calculate this inner product for a given representation. Let us first consider massive spin 1/2 particles in the Dirac representation. From the Lagrangian in \eqn~\eqref{eqn:DiracL}, the canonical momentum is $\pi(x) = i\bar \psi(x) \gamma^0$, and the tensors in the field operator $\widehat{\Psi}_{\nicefrac{1}{2}}$ are the Dirac spinors $u_a(k)$ and $v_a(k)$, where $a = \pm \nicefrac{1}{2}$. Those in the canonical momentum operator are then $i \bar{u}_a(k)$ and $i\bar{v}_a(k)$. 

Given we are interested in the spin vector, we restrict attention to the spin part of the inner product of \eqn~\eqref{eqn:angMomOpInnerProd}. Substituting in the Dirac represenation expressions,
\begin{equation}
\langle p', a|\mathbb{S}_{\mu\nu}^{(1/2)}|p, b\rangle = -\frac{i\hbar}{8E_p} \bar{u}^a(p) \gamma^0 [\gamma_\rho, \gamma_\sigma] u^b(p)\, \del_\Phi(p - p')\,.
\end{equation}
To obtain a simple form for the spin polarisation vector we can combine a variant of the identity in \eqn~\eqref{eqn:3gammaCommutator} with the product $\bar{u}_a(p) \gamma^\mu u_b(p) = 2p^\mu \delta_{ab}$ to obtain
\begin{equation}
\langle p', a|\mathbb{S}_{\mu\nu}^{(1/2)}|p, b\rangle = -\frac{i\hbar}{4E_p}\left(2\eta_{0\rho}p_\sigma \delta_{ab} - 2\eta_{0\sigma}p_\rho \delta_{ab} - i\epsilon_{0\rho\sigma\delta}\bar{u}_a(p) \gamma^\delta \gamma^5 u_b(p)\right) \del_\Phi(p - p')\,.
\end{equation}
Utilising this expression in \eqn~\eqref{eqn:defOfQFTspinVec}, the Levi-Civita tensor eliminates the terms proportional to the 4-momentum, leaving only the pseudovector part. The remaining tensor and gamma product evaluates to $-2E_{p} \bar{u}_a(p) \gamma^\mu \gamma^5 u_b(p)$, so in all we find
\begin{equation}
s^\mu_{ab}(p) = \frac{\hbar}{4m} \bar{u}_a(p) \gamma^\mu \gamma^5 u_b(p)\,,
\end{equation}
as expected. Now let us turn to the massive vector representation. Massive spin 1 vector fields have 3 degrees of freedom, so here the tensors in the field operator are complex polarisation vectors $\varepsilon_i(k)$, where $\varepsilon_i(k)\cdot k = 0$ and $i = 0,\pm 1$. The fields can thus be described as Proca fields, for whom the canonical momenta $\pi^\mu(x) = -\partial_0{B}^\mu(x)$. Thus the constant tensors in ${\Pi}^\mu$ are $-ik^0 \varepsilon^\mu_i(k)$. Classical Proca fields are real, so the inner product of the intrinsic parts of the angular momentum operator is
\begin{multline}
\langle p', i|\mathbb{S}^{(1)}_{\rho\sigma}|p, j\rangle = \frac{\hbar}{2} \langle p, i| \!:\! \sum_{i,j} \int\! \df(k) \left( {\varepsilon_i^*}_\mu(k) (\Sigma_{\rho\sigma})^{\mu}{ }_\nu \varepsilon_j^\nu(k)\, a^\dagger_i(k)\, a_j(k) \right.\\\left. - {\varepsilon_i}_\mu(k) (\Sigma_{\rho\sigma})^{\mu}{ }_\nu {\varepsilon_j^*}^\nu(k)\, a_i(k) \, a^\dagger_j(k) + \cdots\right) \!:\! |p, j\rangle.
\end{multline}
The terms in the ellipsis vanish through the commutation relations, leaving
\begin{equation}
\langle p', i|\mathbb{S}^{(1)}_{\rho\sigma}|p, j\rangle = \hbar\, {\varepsilon^*}^i_\mu(p) (\Sigma_{\rho\sigma})^{\mu\nu} \varepsilon^j_\nu(p)\,\del_\Phi(p - p')\,.
\end{equation}
Using this result in \eqn~\eqref{eqn:defOfQFTspinVec} then immediately leads to the spin polarisation vector quoted in \eqn~\eqref{eqn:vectorSpinVec},
\begin{equation}
s_{ij}^\mu(p) = \frac{i\hbar}{m} \epsilon^{\mu\nu\rho\sigma} p_\nu \varepsilon{^*_i}_\rho(p) {\varepsilon_j}_\sigma(p)\,.
\end{equation}

\section{Spin and scattering observables in electrodynamics}

As an additional application of our formalism we can compute the leading order impulse and angular impulse for spinning particles in classical electrodynamics, whose dynamics are again described by the spin pseudovector in \eqn~\eqref{eqn:GRspinVec}. The equation of motion for a particle with Land\'{e} $g$-factor $g_L$ is the BMT equation \cite{Jackson1999}
\begin{equation}
\frac{ds^\mu}{d\tau} = \frac{e\, g_L}{2m}\left(F^{\mu\nu} s_\nu + u^\mu s_\nu F^{\nu\rho} u_\rho\right) - u^\mu s_\nu \frac{du^\nu}{d\tau}.
\end{equation}
For classical particles $g_L$ takes the universal value $g_L = 2$ \cite{Chung:2018kqs}. The final term is of course given by the Lorentz force, which in this context is modified because the spin introduces a new coupling to the radiation field in the worldline action\footnote{As we aim to compare with results from amplitudes for particles with magnetic dipole factors $g_L$, we must use the classical interaction term related to the magnetic field.},
\begin{equation}
S_\textrm{int} = \frac{e}{m} \int\! d\tau\, \tilde{F}_{\mu\nu}(z(\tau)) u^\mu s^\nu(\tau)\,,
\end{equation}
where the dual field strength $\tilde{F}_{\mu\nu} = -\frac12\epsilon_{\mu\nu\rho\sigma} F^{\rho\sigma}$. Varying with respect to the worldline leads to a modified Lorentz force,
\begin{equation}
\frac{d p^\mu}{d\tau} = e\left(F^{\mu\nu} U_\nu + \frac{d}{d\tau}\left(\tilde{F}^{\mu\nu} \frac{s_\nu}{m}\right) - \frac{s_\rho}{m} u_\nu \partial^\mu \tilde{F}^{\nu\rho}\right)\label{eqn:modLorentz},
\end{equation}
Iteratively solving the Lorentz and BMT equations with straight line trajectories
\begin{equation}
r_1(\tau) = b + u_1\tau, \qquad r_2(\tau) = u_2\tau\,,
\end{equation}
where $u_i$ are now the constant lowest order expansions of the 4-velocities, is enough to then extract the leading order impulse and angular impulse. Iteratively solving the Lorenz gauge Maxwell equation, the radiation field due to particle 2 is given by
\begin{equation}
F_2^{\mu\nu}(x) = ie\int\! \dd^4\wn q\, \del(\wn q\cdot u_2) e^{-i\wn q\cdot x} \frac{\wn q^\mu u_2^\nu - u_2^\mu \wn q^\nu}{\wn q^2}\,.
\end{equation}
Substituting into \eqn~\eqref{eqn:modLorentz}, the modified leading order Lorentz force is then
\begin{multline}
\frac{dp_1^{\mu,(0)}}{d\tau} = ie^2 \int\!\dd^4 \wn q\, \frac{\del(\wn q\cdot u_2)}{\wn q^2} e^{-i\wn q\cdot(b + u_1\tau)} \bigg(\gamma \wn q^\mu - u_2^\mu \wn q\cdot u_1 \\+ \frac{i}{2}(\wn q\cdot u_1)\, \epsilon^{\mu\nu\rho\sigma}\left(\wn q_\rho u_{2\,\sigma} - u_{2\,\rho}\wn q_{\sigma}\right) \frac{s_{1\,\nu}}{m_1} - \frac{i}{2}\wn q^\mu \epsilon_{\nu\alpha\rho\sigma}\left(\wn q^\rho u_2^\sigma - \wn q^\sigma u_2^\rho\right) u_1^\nu \frac{s_1^\alpha}{m_1} \bigg),
\end{multline}
allowing us to obtain the impulse by integrating over the entire domain of $\tau$:
\begin{equation}
\Delta p_1^{\mu,(0)} = ie^2 \int\! \dd^4\wn q\, \del(\wn q\cdot u_1) \del(\wn q\cdot u_2) \frac{e^{-i\wn q\cdot b}}{\wn q^2}\left(\gamma \wn q^\mu + i\wn q^\mu \wn q^\alpha \epsilon_{\alpha\rho\sigma\delta} u_1^\rho u_2^\sigma \frac{s_1^\delta}{m_1} \right).
\end{equation}
The remaining integrals are those defined in \eqn~\eqref{eqn:defOfI}, reducing the expression to
\begin{equation}
\Delta p_1^{\mu,(0)} = -\frac{e^2}{2\pi b^2 \sqrt{\gamma^2 - 1}}\left(\gamma{b}^\mu - \left(\eta^{\mu\alpha} - \frac{2{b}^\mu{b}^\alpha}{b^2}\right) \epsilon_{\alpha\rho\sigma\delta} u_1^\rho u_2^\sigma \frac{s_1^\delta}{m_1}\right).
\end{equation}
This result is a prerequisite for calculating the angular impulse, which similarly integrating the BMT equation over all $\tau$ yields
\begin{equation}
\Delta s_1^{\mu,(0)} = -i\frac{e^2}{m_1} \int\! \dd^4\wn q\, \del(\wn q\cdot u_1)\del(\wn q\cdot u_2) \frac{e^{-i\wn q\cdot b}}{\wn q^2} \epsilon^{\mu\nu\rho\sigma} s_{1\,\nu} u_{1\,\rho} \epsilon_{\sigma\alpha\beta\gamma} \wn q^\alpha u_1^\beta u_2^\gamma - \frac{u_1^\mu s_1^\nu}{m_1} \Delta p_{\nu}^{(0)} \,.
\end{equation}
Once again we reach a form that we can integrate, finding
\begin{multline}
\Delta s_1^{\mu,(0)} = \frac{e^2}{2\pi m_1 b^2 \sqrt{\gamma^2 - 1}}\left(\gamma u_1^\mu s_{1\,\nu} {b}^\nu + \epsilon^{\mu\nu\rho\sigma} s_{1\,\nu} u_{1\,\rho} \epsilon_{\sigma\alpha\beta\gamma} u_1^\alpha u_2^\beta {b}^\gamma \right.\\\left. - u_1^\mu s_{1\,\nu} \left(\eta^{\nu\alpha} - \frac{2{b}^\nu {b}^\alpha}{b^2}\right) \epsilon_{\alpha\rho\sigma\delta} u_1^\rho u_2^\sigma \frac{s_1^\delta}{m_1}\right).\label{eqn:EMAngImp}
\end{multline}

Given our work in gravity, calculating analogous results from amplitudes is trivial. The classical contribution to spin $\nicefrac{1}{2}$-spin 0 scattering in QED can be easily obtained by colour stripping the Yang-Mills amplitude in \eqn~\eqref{eqn:spinorYMamp}, and is
\begin{equation}
\hbar^3 \mathcal{A}_{\textrm{QED}}^{ab} = \frac{4e^2}{\wn{q}^2}\left((p_1 \cdot p_2)\delta^{ab} - \frac{i}{m_1}p_1^\rho \wn q^\sigma p_2^\mu \epsilon_{\rho\sigma\mu\delta} s^{\delta\,ab} + \mathcal{O}(\hbar^2)\right).
\end{equation}
The tensor structures in this amplitude also appear in the gravity amplitudes used in \sect{sec:KerrCalcs}, so the calculations are exactly the same; only prefactors change and we lose higher order spin terms. We find that the linear impulse on particle 1 is
\begin{equation}
\Delta p_1^{\mu,(0)}\big|_{\textrm{QED}} = -\frac{e^2}{2\pi b^2 \sqrt{\gamma^2 - 1}}\left(\gamma {b}^\mu - \left(\eta^{\mu\nu} - \frac{2{b}^\mu{b}^\nu}{b^2}\right) \epsilon_{\nu\rho\sigma\delta} u_1^\rho u_2^\sigma \frac{\spinExp{s^\delta}}{m_1}\right),
\end{equation}
and the angular impulse
\begin{multline}
\Delta s_1^{\mu,(0)}\big|_{\textrm{QED}} = \frac{e^2}{2\pi m_1 b^2 \sqrt{\gamma^2 - 1}}\left(\gamma\, u_1^\mu {b}_\nu \spinExp{s_{1}^\nu} + \epsilon^{\mu\nu\rho\sigma} \spinExp{s_{1\,\nu}} u_{1\,\rho} \epsilon_{\sigma\alpha\beta\gamma} u_1^\alpha u_2^\beta {b}^\gamma \right.\\\left. - u_1^\mu \left(\eta^{\nu\alpha} - \frac{2{b}^\nu {b}^\alpha}{b^2}\right) \epsilon_{\alpha\rho\sigma\tau} u_1^\rho u_2^\sigma \frac{\spinExp{{s_1}_\nu s_1^\tau}}{m_1} \right).
\end{multline}
Comparing with the result obtained from the Lorentz force and BMT equation, we observe an exact match up to the spin expectation values discussed in \sect{sec:discussion}.


\begin{thebibliography}{10}

\bibitem{Hansen1974}
R.~O. Hansen, \emph{Multipole moments of stationary space-times},
  \href{https://doi.org/10.1063/1.1666501}{\emph{Journal of Mathematical
  Physics} {\bfseries 15} (1974) 46--52}.

\bibitem{Ross:2007zza}
A.~Ross and B.~R. Holstein, \emph{{Spin effects in the effective quantum field
  theory of general relativity}},
  \href{https://doi.org/10.1088/1751-8113/40/25/S48}{\emph{J. Phys. A: Math.
  Theor.} {\bfseries 40} (2007) 6973--6978}.

\bibitem{Holstein:2008sx}
B.~R. Holstein and A.~Ross, \emph{{Spin Effects in Long Range Gravitational
  Scattering}},  \href{https://arxiv.org/abs/0802.0716}{{\ttfamily 0802.0716}}.

\bibitem{Vaidya:2014kza}
V.~Vaidya, \emph{{Gravitational spin Hamiltonians from the S matrix}},
  \href{https://doi.org/10.1103/PhysRevD.91.024017}{\emph{Phys. Rev.}
  {\bfseries D91} (2015) 024017},
  [\href{https://arxiv.org/abs/1410.5348}{{\ttfamily 1410.5348}}].

\bibitem{Guevara:2017csg}
A.~Guevara, \emph{{Holomorphic Classical Limit for Spin Effects in
  Gravitational and Electromagnetic Scattering}},
  \href{https://arxiv.org/abs/1706.02314}{{\ttfamily 1706.02314}}.

\bibitem{Arkani-Hamed:2017jhn}
N.~Arkani-Hamed, T.-C. Huang and Y.-t. Huang, \emph{{Scattering Amplitudes For
  All Masses and Spins}},  \href{https://arxiv.org/abs/1709.04891}{{\ttfamily
  1709.04891}}.

\bibitem{Conde:2016vxs} 
  E.~Conde and A.~Marzolla, \emph{{Lorentz Constraints on Massive Three-Point Amplitudes}}, \href{https://doi.org/10.1007/JHEP09(2016)041}{\emph{JHEP} {\bfseries 1609} (2016) 041}, [\href{https://arxiv.org/abs/1601.08113}{{\ttfamily 1601.08113}}].
  
\bibitem{Conde:2016izb} 
  E.~Conde, E.~Joung and K.~Mkrtchyan, \emph{{Spinor-Helicity Three-Point Amplitudes from Local Cubic Interactions}}, \href{https://doi.org/10.1007/JHEP08(2016)040}{\emph{JHEP} {\bfseries 1608} (2016) 040}, [\href{https://arxiv.org/abs/1605.07402}{{\ttfamily 1605.07402}}].

\bibitem{Guevara:2018wpp}
A.~Guevara, A.~Ochirov and J.~Vines, \emph{{Scattering of Spinning Black Holes
  from Exponentiated Soft Factors}},
  \href{https://arxiv.org/abs/1812.06895}{{\ttfamily 1812.06895}}.

\bibitem{Bautista:2019tdr}
Y.~F. Bautista and A.~Guevara, \emph{{From Scattering Amplitudes to Classical
  Physics: Universality, Double Copy and Soft Theorems}},
  \href{https://arxiv.org/abs/1903.12419}{{\ttfamily 1903.12419}}.

\bibitem{Vines:2017hyw}
J.~Vines, \emph{{Scattering of two spinning black holes in post-Minkowskian
  gravity, to all orders in spin, and effective-one-body mappings}},
  \href{https://doi.org/10.1088/1361-6382/aaa3a8}{\emph{Class. Quant. Grav.}
  {\bfseries 35} (2018) 084002},
  [\href{https://arxiv.org/abs/1709.06016}{{\ttfamily 1709.06016}}].

\bibitem{Chung:2018kqs}
M.-Z. Chung, Y.-T. Huang, J.-W. Kim and S.~Lee, \emph{{The simplest massive
  S-matrix: from minimal coupling to Black Holes}}, \href{https://10.1007/JHEP04(2019)15}{\emph{JHEP} {\bfseries 2019} (2019) 156}, 
  [\href{https://arxiv.org/abs/1812.08752}{{\ttfamily 1812.08752}}].

\bibitem{Levi:2015msa}
M.~Levi and J.~Steinhoff, \emph{{Spinning gravitating objects in the effective
  field theory in the post-Newtonian scheme}},
  \href{https://doi.org/10.1007/JHEP09(2015)219}{\emph{JHEP} {\bfseries 09}
  (2015) 219}, [\href{https://arxiv.org/abs/1501.04956}{{\ttfamily
  1501.04956}}].

\bibitem{Abbott:2016blz}
{\scshape LIGO Scientific, Virgo} collaboration, B.~P. Abbott et~al.,
  \emph{{Observation of Gravitational Waves from a Binary Black Hole Merger}},
  \href{https://doi.org/10.1103/PhysRevLett.116.061102}{\emph{Phys. Rev. Lett.}
  {\bfseries 116} (2016) 061102},
  [\href{https://arxiv.org/abs/1602.03837}{{\ttfamily 1602.03837}}].

\bibitem{Buonanno:2014aza}
A.~Buonanno and B.~S. Sathyaprakash, \emph{{Sources of Gravitational Waves:
  Theory and Observations}},  pp.~287--346.
\newblock 2014.
\newblock \href{https://arxiv.org/abs/1410.7832}{{\ttfamily 1410.7832}}.

\bibitem{Buonanno:1998gg}
A.~Buonanno and T.~Damour, \emph{{Effective one-body approach to general
  relativistic two-body dynamics}},
  \href{https://doi.org/10.1103/PhysRevD.59.084006}{\emph{Phys. Rev.}
  {\bfseries D59} (1999) 084006},
  [\href{https://arxiv.org/abs/gr-qc/9811091}{{\ttfamily gr-qc/9811091}}].

\bibitem{Buonanno:2000ef}
A.~Buonanno and T.~Damour, \emph{{Transition from inspiral to plunge in binary
  black hole coalescences}},
  \href{https://doi.org/10.1103/PhysRevD.62.064015}{\emph{Phys. Rev.}
  {\bfseries D62} (2000) 064015},
  [\href{https://arxiv.org/abs/gr-qc/0001013}{{\ttfamily gr-qc/0001013}}].

\bibitem{Damour:2001tu}
T.~Damour, \emph{{Coalescence of two spinning black holes: an effective
  one-body approach}},
  \href{https://doi.org/10.1103/PhysRevD.64.124013}{\emph{Phys. Rev.}
  {\bfseries D64} (2001) 124013},
  [\href{https://arxiv.org/abs/gr-qc/0103018}{{\ttfamily gr-qc/0103018}}].

\bibitem{Damour:2008qf}
T.~Damour, P.~Jaranowski and G.~Schaefer, \emph{{Effective one body approach to
  the dynamics of two spinning black holes with next-to-leading order
  spin-orbit coupling}},
  \href{https://doi.org/10.1103/PhysRevD.78.024009}{\emph{Phys. Rev.}
  {\bfseries D78} (2008) 024009},
  [\href{https://arxiv.org/abs/0803.0915}{{\ttfamily 0803.0915}}].

\bibitem{Barausse:2009aa}
E.~Barausse, E.~Racine and A.~Buonanno, \emph{{Hamiltonian of a spinning
  test-particle in curved spacetime}},
  \href{https://doi.org/10.1103/PhysRevD.85.069904,
  10.1103/PhysRevD.80.104025}{\emph{Phys. Rev.} {\bfseries D80} (2009) 104025},
  [\href{https://arxiv.org/abs/0907.4745}{{\ttfamily 0907.4745}}].

\bibitem{Barausse:2009xi}
E.~Barausse and A.~Buonanno, \emph{{An Improved effective-one-body Hamiltonian
  for spinning black-hole binaries}},
  \href{https://doi.org/10.1103/PhysRevD.81.084024}{\emph{Phys. Rev.}
  {\bfseries D81} (2010) 084024},
  [\href{https://arxiv.org/abs/0912.3517}{{\ttfamily 0912.3517}}].

\bibitem{Barausse:2011ys}
E.~Barausse and A.~Buonanno, \emph{{Extending the effective-one-body
  Hamiltonian of black-hole binaries to include next-to-next-to-leading
  spin-orbit couplings}},
  \href{https://doi.org/10.1103/PhysRevD.84.104027}{\emph{Phys. Rev.}
  {\bfseries D84} (2011) 104027},
  [\href{https://arxiv.org/abs/1107.2904}{{\ttfamily 1107.2904}}].

\bibitem{Damour:2014sva}
T.~Damour and A.~Nagar, \emph{{New effective-one-body description of coalescing
  nonprecessing spinning black-hole binaries}},
  \href{https://doi.org/10.1103/PhysRevD.90.044018}{\emph{Phys. Rev.}
  {\bfseries D90} (2014) 044018},
  [\href{https://arxiv.org/abs/1406.6913}{{\ttfamily 1406.6913}}].

\bibitem{Bini:2017wfr}
D.~Bini and T.~Damour, \emph{{Gravitational scattering of two black holes at
  the fourth post-Newtonian approximation}},
  \href{https://doi.org/10.1103/PhysRevD.96.064021}{\emph{Phys. Rev.}
  {\bfseries D96} (2017) 064021},
  [\href{https://arxiv.org/abs/1706.06877}{{\ttfamily 1706.06877}}].

\bibitem{Bini:2017xzy}
D.~Bini and T.~Damour, \emph{{Gravitational spin-orbit coupling in binary
  systems, post-Minkowskian approximation and effective one-body theory}},
  \href{https://doi.org/10.1103/PhysRevD.96.104038}{\emph{Phys. Rev.}
  {\bfseries D96} (2017) 104038},
  [\href{https://arxiv.org/abs/1709.00590}{{\ttfamily 1709.00590}}].

\bibitem{Bini:2018ywr}
D.~Bini and T.~Damour, \emph{{Gravitational spin-orbit coupling in binary
  systems at the second post-Minkowskian approximation}},
  \href{https://doi.org/10.1103/PhysRevD.98.044036}{\emph{Phys. Rev.}
  {\bfseries D98} (2018) 044036},
  [\href{https://arxiv.org/abs/1805.10809}{{\ttfamily 1805.10809}}].

\bibitem{Vines:2018gqi}
J.~Vines, J.~Steinhoff and A.~Buonanno, \emph{{Spinning-black-hole scattering
  and the test-black-hole limit at second post-Minkowskian order}},
  \href{https://doi.org/10.1103/PhysRevD.99.064054}{\emph{Phys. Rev.}
  {\bfseries D99} (2019) 064054},
  [\href{https://arxiv.org/abs/1812.00956}{{\ttfamily 1812.00956}}].

\bibitem{Goldberger:2004jt}
W.~D. Goldberger and I.~Z. Rothstein, \emph{{An Effective Field Theory of
  Gravity for Extended Objects}},
  \href{https://doi.org/10.1103/PhysRevD.73.104029}{\emph{Phys. Rev.}
  {\bfseries D73} (2006) 104029},
  [\href{https://arxiv.org/abs/hep-th/0409156}{{\ttfamily hep-th/0409156}}].

\bibitem{Porto:2005ac}
R.~A. Porto, \emph{{Post-Newtonian corrections to the motion of spinning bodies
  in NRGR}}, \href{https://doi.org/10.1103/PhysRevD.73.104031}{\emph{Phys.
  Rev.} {\bfseries D73} (2006) 104031},
  [\href{https://arxiv.org/abs/gr-qc/0511061}{{\ttfamily gr-qc/0511061}}].

\bibitem{Porto:2006bt}
R.~A. Porto and I.~Z. Rothstein, \emph{{The Hyperfine Einstein-Infeld-Hoffmann
  potential}}, \href{https://doi.org/10.1103/PhysRevLett.97.021101}{\emph{Phys.
  Rev. Lett.} {\bfseries 97} (2006) 021101},
  [\href{https://arxiv.org/abs/gr-qc/0604099}{{\ttfamily gr-qc/0604099}}].

\bibitem{Porto:2008tb}
R.~A. Porto and I.~Z. Rothstein, \emph{{Spin(1)Spin(2) Effects in the Motion of
  Inspiralling Compact Binaries at Third Order in the Post-Newtonian
  Expansion}}, \href{https://doi.org/10.1103/PhysRevD.78.044012,
  10.1103/PhysRevD.81.029904}{\emph{Phys. Rev.} {\bfseries D78} (2008) 044012},
  [\href{https://arxiv.org/abs/0802.0720}{{\ttfamily 0802.0720}}].

\bibitem{Levi:2011eq}
M.~Levi, \emph{{Binary dynamics from spin1-spin2 coupling at fourth
  post-Newtonian order}},
  \href{https://doi.org/10.1103/PhysRevD.85.064043}{\emph{Phys. Rev.}
  {\bfseries D85} (2012) 064043},
  [\href{https://arxiv.org/abs/1107.4322}{{\ttfamily 1107.4322}}].

\bibitem{Levi:2016ofk}
M.~Levi and J.~Steinhoff, \emph{{Complete conservative dynamics for
  inspiralling compact binaries with spins at fourth post-Newtonian order}},
  \href{https://arxiv.org/abs/1607.04252}{{\ttfamily 1607.04252}}.

\bibitem{Porto:2016pyg}
R.~A. Porto, \emph{{The effective field theorist’s approach to gravitational
  dynamics}}, \href{https://doi.org/10.1016/j.physrep.2016.04.003}{\emph{Phys.
  Rept.} {\bfseries 633} (2016) 1--104},
  [\href{https://arxiv.org/abs/1601.04914}{{\ttfamily 1601.04914}}].

\bibitem{Levi:2018nxp}
M.~Levi, \emph{{Effective Field Theories of Post-Newtonian Gravity: A
  comprehensive review}},  \href{https://arxiv.org/abs/1807.01699}{{\ttfamily
  1807.01699}}.

\bibitem{Vines:2016qwa}
J.~Vines and J.~Steinhoff, \emph{{Spin-multipole effects in binary black holes
  and the test-body limit}},
  \href{https://doi.org/10.1103/PhysRevD.97.064010}{\emph{Phys. Rev.}
  {\bfseries D97} (2018) 064010},
  [\href{https://arxiv.org/abs/1606.08832}{{\ttfamily 1606.08832}}].

\bibitem{Siemonsen:2017yux}
N.~Siemonsen, J.~Steinhoff and J.~Vines, \emph{{Gravitational waves from
  spinning binary black holes at the leading post-Newtonian orders at all
  orders in spin}},
  \href{https://doi.org/10.1103/PhysRevD.97.124046}{\emph{Phys. Rev.}
  {\bfseries D97} (2018) 124046},
  [\href{https://arxiv.org/abs/1712.08603}{{\ttfamily 1712.08603}}].

\bibitem{Iwasaki:1971}
Y.~Iwasaki, \emph{{Quantum Theory of Gravitation vs. Classical Theory:
  Fourth-Order Potential}},
  \href{https://doi.org/10.1143/PTP.46.1587}{\emph{Prog. Theor. Phys.}
  {\bfseries 46} (1971) 1587--1609}.

\bibitem{Duff:1973zz}
M.~J. Duff, \emph{{Quantum Tree Graphs and the Schwarzschild Solution}},
  \href{https://doi.org/10.1103/PhysRevD.7.2317}{\emph{Phys. Rev.} {\bfseries
  D7} (1973) 2317--2326}.

\bibitem{Donoghue:1993eb}
J.~F. Donoghue, \emph{{Leading quantum correction to the Newtonian potential}},
  \href{https://doi.org/10.1103/PhysRevLett.72.2996}{\emph{Phys. Rev. Lett.}
  {\bfseries 72} (1994) 2996--2999},
  [\href{https://arxiv.org/abs/gr-qc/9310024}{{\ttfamily gr-qc/9310024}}].

\bibitem{Donoghue:1994dn}
J.~F. Donoghue, \emph{{General relativity as an effective field theory: The
  leading quantum corrections}},
  \href{https://doi.org/10.1103/PhysRevD.50.3874}{\emph{Phys. Rev.} {\bfseries
  D50} (1994) 3874--3888},
  [\href{https://arxiv.org/abs/gr-qc/9405057}{{\ttfamily gr-qc/9405057}}].

\bibitem{Bjerrum-Bohr:2002kt}
N.~E.~J. Bjerrum-Bohr, J.~F. Donoghue and B.~R. Holstein, \emph{{Quantum
  gravitational corrections to the nonrelativistic scattering potential of two
  masses}}, \href{https://doi.org/10.1103/PhysRevD.71.069903,
  10.1103/PhysRevD.67.084033}{\emph{Phys. Rev.} {\bfseries D67} (2003) 084033},
  [\href{https://arxiv.org/abs/hep-th/0211072}{{\ttfamily hep-th/0211072}}].

\bibitem{Khriplovich:2004cx}
I.~B. Khriplovich and G.~G. Kirilin, \emph{{Quantum long range interactions in
  general relativity}}, \href{https://doi.org/10.1134/1.1777618}{\emph{J. Exp.
  Theor. Phys.} {\bfseries 98} (2004) 1063--1072},
  [\href{https://arxiv.org/abs/gr-qc/0402018}{{\ttfamily gr-qc/0402018}}].

\bibitem{Holstein:2004dn}
B.~R. Holstein and J.~F. Donoghue, \emph{{Classical Physics and Quantum
  Loops}}, \href{https://doi.org/10.1103/PhysRevLett.93.201602}{\emph{Phys.
  Rev. Lett.} {\bfseries 93} (2004) 201602},
  [\href{https://arxiv.org/abs/hep-th/0405239}{{\ttfamily hep-th/0405239}}].

\bibitem{Neill:2013wsa}
D.~Neill and I.~Z. Rothstein, \emph{{Classical Space-Times from the S Matrix}},
  \href{https://doi.org/10.1016/j.nuclphysb.2013.09.007}{\emph{Nucl. Phys.}
  {\bfseries B877} (2013) 177--189},
  [\href{https://arxiv.org/abs/1304.7263}{{\ttfamily 1304.7263}}].

\bibitem{Bjerrum-Bohr:2013bxa}
N.~E.~J. Bjerrum-Bohr, J.~F. Donoghue and P.~Vanhove, \emph{{On-shell
  Techniques and Universal Results in Quantum Gravity}},
  \href{https://doi.org/10.1007/JHEP02(2014)111}{\emph{JHEP} {\bfseries 02}
  (2014) 111}, [\href{https://arxiv.org/abs/1309.0804}{{\ttfamily 1309.0804}}].

\bibitem{Bjerrum-Bohr:2014lea}
N.~E.~J. Bjerrum-Bohr, B.~R. Holstein, L.~Plant\'{e} and P.~Vanhove,
  \emph{{Graviton-Photon Scattering}},
  \href{https://doi.org/10.1103/PhysRevD.91.064008}{\emph{Phys. Rev.}
  {\bfseries D91} (2015) 064008},
  [\href{https://arxiv.org/abs/1410.4148}{{\ttfamily 1410.4148}}].

\bibitem{Bjerrum-Bohr:2014zsa}
N.~E.~J. Bjerrum-Bohr, J.~F. Donoghue, B.~R. Holstein, L.~Plant\'{e} and
  P.~Vanhove, \emph{{Bending of Light in Quantum Gravity}},
  \href{https://doi.org/10.1103/PhysRevLett.114.061301}{\emph{Phys. Rev. Lett.}
  {\bfseries 114} (2015) 061301},
  [\href{https://arxiv.org/abs/1410.7590}{{\ttfamily 1410.7590}}].

\bibitem{Bjerrum-Bohr:2016hpa}
N.~E.~J. Bjerrum-Bohr, J.~F. Donoghue, B.~R. Holstein, L.~Plant\'{e} and
  P.~Vanhove, \emph{{Light-like Scattering in Quantum Gravity}},
  \href{https://doi.org/10.1007/JHEP11(2016)117}{\emph{JHEP} {\bfseries 11}
  (2016) 117}, [\href{https://arxiv.org/abs/1609.07477}{{\ttfamily
  1609.07477}}].

\bibitem{Bjerrum-Bohr:2017dxw}
N.~E.~J. Bjerrum-Bohr, B.~R. Holstein, J.~F. Donoghue, L.~Plant\'{e} and
  P.~Vanhove, \emph{{Illuminating Light Bending}},
  \href{https://doi.org/10.22323/1.292.0077}{\emph{PoS} {\bfseries CORFU2016}
  (2017) 077}, [\href{https://arxiv.org/abs/1704.01624}{{\ttfamily
  1704.01624}}].

\bibitem{Cachazo:2017jef}
F.~Cachazo and A.~Guevara, \emph{{Leading Singularities and Classical
  Gravitational Scattering}},
  \href{https://arxiv.org/abs/1705.10262}{{\ttfamily 1705.10262}}.

\bibitem{Damour:2016gwp}
T.~Damour, \emph{{Gravitational scattering, post-Minkowskian approximation and
  Effective One-Body theory}},
  \href{https://doi.org/10.1103/PhysRevD.94.104015}{\emph{Phys. Rev.}
  {\bfseries D94} (2016) 104015},
  [\href{https://arxiv.org/abs/1609.00354}{{\ttfamily 1609.00354}}].

\bibitem{Damour:2017zjx}
T.~Damour, \emph{{High-energy gravitational scattering and the general
  relativistic two-body problem}},
  \href{https://doi.org/10.1103/PhysRevD.97.044038}{\emph{Phys. Rev.}
  {\bfseries D97} (2018) 044038},
  [\href{https://arxiv.org/abs/1710.10599}{{\ttfamily 1710.10599}}].

\bibitem{Bjerrum-Bohr:2018xdl}
N.~E.~J. Bjerrum-Bohr, P.~H. Damgaard, G.~Festuccia, L.~Plant\'{e} and
  P.~Vanhove, \emph{{General Relativity from Scattering Amplitudes}},
  \href{https://doi.org/10.1103/PhysRevLett.121.171601}{\emph{Phys. Rev. Lett.}
  {\bfseries 121} (2018) 171601},
  [\href{https://arxiv.org/abs/1806.04920}{{\ttfamily 1806.04920}}].

\bibitem{Cheung:2018wkq}
C.~Cheung, I.~Z. Rothstein and M.~P. Solon, \emph{{From Scattering Amplitudes
  to Classical Potentials in the Post-Minkowskian Expansion}},
  \href{https://doi.org/10.1103/PhysRevLett.121.251101}{\emph{Phys. Rev. Lett.}
  {\bfseries 121} (2018) 251101},
  [\href{https://arxiv.org/abs/1808.02489}{{\ttfamily 1808.02489}}].

\bibitem{Bern:2019nnu}
Z.~Bern, C.~Cheung, R.~Roiban, C.-H. Shen, M.~P. Solon and M.~Zeng,
  \emph{{Scattering Amplitudes and the Conservative Hamiltonian for Binary
  Systems at Third Post-Minkowskian Order}}, \href{https://doi.org/10.1103/PhysRevLett.122.201603}{\emph{Phys. Rev. Lett.} {\bfseries 122} (2019) 201603},
  [\href{https://arxiv.org/abs/1901.04424}{{\ttfamily 1901.04424}}].

\bibitem{Cristofoli:2019neg}
A.~Cristofoli, N.~E.~J. Bjerrum-Bohr, P.~H. Damgaard and P.~Vanhove, \emph{{On
  Post-Minkowskian Hamiltonians in General Relativity}},
  \href{https://arxiv.org/abs/1906.01579}{{\ttfamily 1906.01579}}.

\bibitem{Bern:2008qj}
Z.~Bern, J.~J.~M. Carrasco and H.~Johansson, \emph{{New Relations for
  Gauge-Theory Amplitudes}},
  \href{https://doi.org/10.1103/PhysRevD.78.085011}{\emph{Phys. Rev.}
  {\bfseries D78} (2008) 085011},
  [\href{https://arxiv.org/abs/0805.3993}{{\ttfamily 0805.3993}}].

\bibitem{Bern:2010ue}
Z.~Bern, J.~J.~M. Carrasco and H.~Johansson, \emph{{Perturbative Quantum
  Gravity as a Double Copy of Gauge Theory}},
  \href{https://doi.org/10.1103/PhysRevLett.105.061602}{\emph{Phys. Rev. Lett.}
  {\bfseries 105} (2010) 061602},
  [\href{https://arxiv.org/abs/1004.0476}{{\ttfamily 1004.0476}}].

\bibitem{Bern:2010yg}
Z.~Bern, T.~Dennen, Y.-t. Huang and M.~Kiermaier, \emph{{Gravity as the Square
  of Gauge Theory}},
  \href{https://doi.org/10.1103/PhysRevD.82.065003}{\emph{Phys. Rev.}
  {\bfseries D82} (2010) 065003},
  [\href{https://arxiv.org/abs/1004.0693}{{\ttfamily 1004.0693}}].

\bibitem{Johansson:2014zca}
H.~Johansson and A.~Ochirov, \emph{{Pure Gravities via Color-Kinematics Duality
  for Fundamental Matter}},
  \href{https://doi.org/10.1007/JHEP11(2015)046}{\emph{JHEP} {\bfseries 11}
  (2015) 046}, [\href{https://arxiv.org/abs/1407.4772}{{\ttfamily 1407.4772}}].

\bibitem{Johansson:2015oia}
H.~Johansson and A.~Ochirov, \emph{{Color-Kinematics Duality for QCD
  Amplitudes}}, \href{https://doi.org/10.1007/JHEP01(2016)170}{\emph{JHEP}
  {\bfseries 01} (2016) 170},
  [\href{https://arxiv.org/abs/1507.00332}{{\ttfamily 1507.00332}}].

\bibitem{Carrasco:2015iwa}
J.~J.~M. Carrasco, \emph{{Gauge and Gravity Amplitude Relations}},  in
  \emph{{Proceedings, Theoretical Advanced Study Institute in Elementary
  Particle Physics: Journeys Through the Precision Frontier: Amplitudes for
  Colliders (TASI 2014): Boulder, Colorado, June 2-27, 2014}}, pp.~477--557,
  WSP, WSP, 2015, \href{https://arxiv.org/abs/1506.00974}{{\ttfamily
  1506.00974}}, \href{https://doi.org/10.1142/9789814678766_0011}{DOI}.

\bibitem{Bern:2018jmv}
Z.~Bern, J.~J. Carrasco, W.-M. Chen, A.~Edison, H.~Johansson, J.~Parra-Martinez
  et~al., \emph{{Ultraviolet Properties of $\mathcal N = 8$ Supergravity at
  Five Loops}}, \href{https://doi.org/10.1103/PhysRevD.98.086021}{\emph{Phys.
  Rev.} {\bfseries D98} (2018) 086021},
  [\href{https://arxiv.org/abs/1804.09311}{{\ttfamily 1804.09311}}].

\bibitem{Bern:2019isl}
Z.~Bern, D.~Kosower and J.~Parra-Martinez, \emph{{Two-loop n-point anomalous
  amplitudes in $\mathcal N=4$ supergravity}},
  \href{https://arxiv.org/abs/1905.05151}{{\ttfamily 1905.05151}}.

\bibitem{Monteiro:2014cda}
R.~Monteiro, D.~O'Connell and C.~D. White, \emph{{Black holes and the double
  copy}}, \href{https://doi.org/10.1007/JHEP12(2014)056}{\emph{JHEP} {\bfseries
  12} (2014) 056}, [\href{https://arxiv.org/abs/1410.0239}{{\ttfamily
  1410.0239}}].

\bibitem{Luna:2015paa}
A.~Luna, R.~Monteiro, D.~O'Connell and C.~D. White, \emph{{The classical double
  copy for Taub–NUT spacetime}},
  \href{https://doi.org/10.1016/j.physletb.2015.09.021}{\emph{Phys. Lett.}
  {\bfseries B750} (2015) 272--277},
  [\href{https://arxiv.org/abs/1507.01869}{{\ttfamily 1507.01869}}].

\bibitem{Luna:2016due}
A.~Luna, R.~Monteiro, I.~Nicholson, D.~O'Connell and C.~D. White, \emph{{The
  double copy: Bremsstrahlung and accelerating black holes}},
  \href{https://doi.org/10.1007/JHEP06(2016)023}{\emph{JHEP} {\bfseries 06}
  (2016) 023}, [\href{https://arxiv.org/abs/1603.05737}{{\ttfamily
  1603.05737}}].

\bibitem{Goldberger:2016iau}
W.~D. Goldberger and A.~K. Ridgway, \emph{{Radiation and the classical double
  copy for color charges}},
  \href{https://doi.org/10.1103/PhysRevD.95.125010}{\emph{Phys. Rev.}
  {\bfseries D95} (2017) 125010},
  [\href{https://arxiv.org/abs/1611.03493}{{\ttfamily 1611.03493}}].

\bibitem{Luna:2016hge}
A.~Luna, R.~Monteiro, I.~Nicholson, A.~Ochirov, D.~O'Connell, N.~Westerberg
  et~al., \emph{{Perturbative spacetimes from Yang-Mills theory}},
  \href{https://doi.org/10.1007/JHEP04(2017)069}{\emph{JHEP} {\bfseries 04}
  (2017) 069}, [\href{https://arxiv.org/abs/1611.07508}{{\ttfamily
  1611.07508}}].

\bibitem{Adamo:2017nia}
T.~Adamo, E.~Casali, L.~Mason and S.~Nekovar, \emph{{Scattering on plane waves
  and the double copy}},
  \href{https://doi.org/10.1088/1361-6382/aa9961}{\emph{Class. Quant. Grav.}
  {\bfseries 35} (2018) 015004},
  [\href{https://arxiv.org/abs/1706.08925}{{\ttfamily 1706.08925}}].

\bibitem{Goldberger:2017frp}
W.~D. Goldberger, S.~G. Prabhu and J.~O. Thompson, \emph{{Classical gluon and
  graviton radiation from the bi-adjoint scalar double copy}},
  \href{https://doi.org/10.1103/PhysRevD.96.065009}{\emph{Phys. Rev.}
  {\bfseries D96} (2017) 065009},
  [\href{https://arxiv.org/abs/1705.09263}{{\ttfamily 1705.09263}}].

\bibitem{Bahjat-Abbas:2017htu}
N.~Bahjat-Abbas, A.~Luna and C.~D. White, \emph{{The Kerr-Schild double copy in
  curved spacetime}},
  \href{https://doi.org/10.1007/JHEP12(2017)004}{\emph{JHEP} {\bfseries 12}
  (2017) 004}, [\href{https://arxiv.org/abs/1710.01953}{{\ttfamily
  1710.01953}}].

\bibitem{Carrillo-Gonzalez:2017iyj}
M.~Carrillo-González, R.~Penco and M.~Trodden, \emph{{The classical double
  copy in maximally symmetric spacetimes}},
  \href{https://doi.org/10.1007/JHEP04(2018)028}{\emph{JHEP} {\bfseries 04}
  (2018) 028}, [\href{https://arxiv.org/abs/1711.01296}{{\ttfamily
  1711.01296}}].

\bibitem{Goldberger:2017vcg}
W.~D. Goldberger and A.~K. Ridgway, \emph{{Bound states and the classical
  double copy}}, \href{https://doi.org/10.1103/PhysRevD.97.085019}{\emph{Phys.
  Rev.} {\bfseries D97} (2018) 085019},
  [\href{https://arxiv.org/abs/1711.09493}{{\ttfamily 1711.09493}}].

\bibitem{Shen:2018ebu}
C.-H. Shen, \emph{{Gravitational Radiation from Color-Kinematics Duality}},
  \href{https://doi.org/10.1007/JHEP11(2018)162}{\emph{JHEP} {\bfseries 11}
  (2018) 162}, [\href{https://arxiv.org/abs/1806.07388}{{\ttfamily
  1806.07388}}].

\bibitem{Plefka:2018dpa}
J.~Plefka, J.~Steinhoff and W.~Wormsbecher, \emph{{Effective action of dilaton
  gravity as the classical double copy of Yang-Mills theory}},
  \href{https://doi.org/10.1103/PhysRevD.99.024021}{\emph{Phys. Rev.}
  {\bfseries D99} (2019) 024021},
  [\href{https://arxiv.org/abs/1807.09859}{{\ttfamily 1807.09859}}].

\bibitem{Berman:2018hwd}
D.~S. Berman, E.~Chacón, A.~Luna and C.~D. White, \emph{{The self-dual
  classical double copy, and the Eguchi-Hanson instanton}},
  \href{https://doi.org/10.1007/JHEP01(2019)107}{\emph{JHEP} {\bfseries 01}
  (2019) 107}, [\href{https://arxiv.org/abs/1809.04063}{{\ttfamily
  1809.04063}}].

\bibitem{CarrilloGonzalez:2019gof}
M.~Carrillo~Gonz\'{a}lez, B.~Melcher, K.~Ratliff, S.~Watson and C.~D. White,
  \emph{{The classical double copy in three spacetime dimensions}},
  \href{https://arxiv.org/abs/1904.11001}{{\ttfamily 1904.11001}}.

\bibitem{Luna:2018dpt}
A.~Luna, R.~Monteiro, I.~Nicholson and D.~O'Connell, \emph{{Type D Spacetimes
  and the Weyl Double Copy}}, \href{https://doi.org/10.1088/1361-6382/ab03e6}{\emph{Class. Quant. Grav.} {\bfseries 36} (2019) 6}, 
  [\href{https://arxiv.org/abs/1810.08183}{{\ttfamily 1810.08183}}].

\bibitem{Goldberger:2017ogt}
W.~D. Goldberger, J.~Li and S.~G. Prabhu, \emph{{Spinning particles, axion
  radiation, and the classical double copy}},
  \href{https://doi.org/10.1103/PhysRevD.97.105018}{\emph{Phys. Rev.}
  {\bfseries D97} (2018) 105018},
  [\href{https://arxiv.org/abs/1712.09250}{{\ttfamily 1712.09250}}].

\bibitem{Li:2018qap}
J.~Li and S.~G. Prabhu, \emph{{Gravitational radiation from the classical
  spinning double copy}},
  \href{https://doi.org/10.1103/PhysRevD.97.105019}{\emph{Phys. Rev.}
  {\bfseries D97} (2018) 105019},
  [\href{https://arxiv.org/abs/1803.02405}{{\ttfamily 1803.02405}}].

\bibitem{Antonelli:2019ytb}
A.~Antonelli, A.~Buonanno, J.~Steinhoff, M.~van~de Meent and J.~Vines,
  \emph{{Energetics of two-body Hamiltonians in post-Minkowskian gravity}}, \href{https:// 	10.1103/PhysRevD.100.045003}{\emph{Phys. Rev.} {\bfseries D100} (2019) 045003},
  [\href{https://arxiv.org/abs/1901.07102}{{\ttfamily 1901.07102}}].

\bibitem{Kosower:2018adc}
D.~A. Kosower, B.~Maybee and D.~O'Connell, \emph{{Amplitudes, Observables, and
  Classical Scattering}},
  \href{https://doi.org/10.1007/JHEP02(2019)137}{\emph{JHEP} {\bfseries 02}
  (2019) 137}, [\href{https://arxiv.org/abs/1811.10950}{{\ttfamily
  1811.10950}}].

\bibitem{Luna:2017dtq}
A.~Luna, I.~Nicholson, D.~O'Connell and C.~D. White, \emph{{Inelastic Black
  Hole Scattering from Charged Scalar Amplitudes}},
  \href{https://doi.org/10.1007/JHEP03(2018)044}{\emph{JHEP} {\bfseries 03}
  (2018) 044}, [\href{https://arxiv.org/abs/1711.03901}{{\ttfamily
  1711.03901}}].

\bibitem{Fokker:1929}
A.~D. Fokker, \emph{{Relativiteitstheorie}}.
\newblock P. Noordhoff, 1929.

\bibitem{Tulczyjew:1959}
W.~M. Tulczyjew, \emph{{Motion of multipole particles in general relativity
  theory}}, \href{https://doi.org/10.3847/2041-8213/ab0ec7}{\emph{Acta Phys.
  Polon.} {\bfseries 18} (1959) 393}.

\bibitem{Mathisson:1937zz}
M.~Mathisson, \emph{{Neue mechanik materieller systemes}}, {\emph{Acta Phys.
  Polon.} {\bfseries 6} (1937) 163--2900}.

\bibitem{Mathisson:2010}
M.~Mathisson, \emph{{Republication of: New mechanics of material systems}},
  \href{https://doi.org/https://doi.org/10.1007/s10714-010-0939-y}{\emph{Gen.
  Relativ. Grav.} {\bfseries 42} (2010) 1011}.

\bibitem{Papapetrou:1951pa}
A.~Papapetrou, \emph{{Spinning test particles in general relativity. 1.}},
  \href{https://doi.org/10.1098/rspa.1951.0200}{\emph{Proc. Roy. Soc. Lond.}
  {\bfseries A209} (1951) 248--258}.

\bibitem{Dixon1979}
W.~G. Dixon, \emph{{A covariant multipole formalism for extended test bodies in
  general relativity}},  in \emph{{Proceedings of the International School of
  Physics Enrico Fermi LXVII}} (J.~Ehlers, ed.), pp.~156 -- 219, North Holland,
  1979.

\bibitem{Dixon:2015vxa}
W.~G. Dixon, \emph{{The New Mechanics of Myron Mathisson and Its Subsequent
  Development}}, \href{https://doi.org/10.1007/978-3-319-18335-0_1}{\emph{Fund.
  Theor. Phys.} {\bfseries 179} (2015) 1--66}.

\bibitem{Weinberg:1972kfs}
S.~Weinberg, \emph{Gravitation and Cosmology: Principles and Applications of
  the General Theory of Relativity}.
\newblock Wiley, 1972.

\bibitem{Cotogno:2019xcl} 
S.~Cotogno, C.~Lorc\'e and P.~Lowdon, \emph{{Poincar\'e constraints on the gravitational form factors for massive states with arbitrary spin}},
\href{https://doi.org/10.1103/PhysRevD.100.045003}{\emph{Phys. Rev.} {\bfseries D100}
	(2019) 045003}, [\href{https://arxiv.org/abs/1905.11969}{{\ttfamily
		1905.11969}}].

\bibitem{Lorce:2019sbq} 
C.~Lorc\'e and P.~Lowdon, \emph{{Universality of the Poincar\'e gravitational form factor constraints}},
\href{https://arxiv.org/abs/1908.02567}{{\ttfamily 1908.02567}}.

  
\bibitem{Lorce:2017isp} 
C.~Lorc\'e, \emph{{New explicit expressions for Dirac bilinears}},
\href{https://doi.org/10.1103/PhysRevD.97.016005}{\emph{Phys. Rev.} {\bfseries D97}
	(2018) 016005}, [\href{https://arxiv.org/abs/1705.08370}{{\ttfamily
		1705.08370}}].

\bibitem{Bogoliubov:1980}
N.~N. Bogoliubov and D.~V. Shirkov, \emph{{Introduction to the theory of
  quantized fields}}.
\newblock Wiley, 1980.

\bibitem{Weinberg:1995mt}
S.~Weinberg, \emph{{The Quantum Theory of Fields. Vol. 1: Foundations}}.
\newblock Cambridge University Press, 2005.


\bibitem{Jackson1999}
J.~D. Jackson, \emph{Classical Electrodynamics}.
\newblock Wiley, 1999.

\end{thebibliography}

\providecommand{\href}[2]{#2}\begingroup\raggedright\endgroup

\end{document}